\documentclass[twocolumn,eqsecnum,aps,superscriptaddress]{revtex4}

\usepackage{graphicx}%
\usepackage{dcolumn}
\usepackage{amsmath}
\usepackage{amssymb}
\usepackage{multirow}

\newcommand {\AGeV}     {\mbox{A$ \cdot$GeV/c}}
\newcommand {\ntma}     {\mbox{N$_{\text{TMA}}$}}
\newcommand {\ezcal}    {\mbox{E$_{\text{ZCAL}}$}}
\newcommand {\enucl}    {\mbox{E$_{\text{nucleon}}$}}

\newcommand {\ppbar}    {\mbox{$p\overline{p}$}}
\newcommand {\ZCNproj}  {\mbox{$^{ZC}N_{\text{proj}}$}}
\newcommand {\Nproj}    {\mbox{$N_{\text{proj}}$}}

\newcommand {\Npart}    {\mbox{$N_{\text{total}}$}}

\newcommand {\Rproj}    {\mbox{$N_{\text{proj}}^{1/3}$}}
\newcommand {\Rtarg}    {\mbox{$N_{\text{targ}}^{1/3}$}}
\newcommand {\Rpart}    {\mbox{$N_{\text{total}}^{1/3}$}}
\newcommand {\ZCRproj}  {\mbox{$^{ZC}N_{\text{proj}}^{1/3}$}}
\newcommand {\chisq}    {\mbox{$\chi^2$}}
\newcommand {\Qinv}     {\mbox{$Q_{\text{inv}}$}}
\newcommand {\Rinv}     {\mbox{$R_{\text{inv}}$}}

\newcommand {\Rs}       {\mbox{$R_{\text s}$}}
\newcommand {\Ro}       {\mbox{$R_{\text o}$}}
\newcommand {\Rl}       {\mbox{$R_{\text l}$}}
\newcommand {\Rlo}      {\mbox{$R_{\text{lo}}$}}
\newcommand {\myt}      {\mbox{$\tau$}}

\newcommand {\qs}       {\mbox{$q_{\text s}$}}
\newcommand {\qo}       {\mbox{$q_{\text o}$}}
\newcommand {\ql}       {\mbox{$q_{\text l}$}}

\newcommand {\qzero}    {\mbox{$q_{0}$}}
\newcommand {\kT}       {\mbox{$k_T$}}
\newcommand {\mT}       {\mbox{$m_T$}}

\newcommand {\mmT}      {\mbox{$<m_T>$}}
\newcommand {\mmTs}     {\mbox{$<m^*_T>$}}

\newcommand {\dg}       {\mbox{$^\circ$}}

\newcommand {\bsrc}     {\mbox{$\beta_{\text{s}}$}}

\newcommand {\Yfit}     {\mbox{$Y_{\text{fit}}$}}
\newcommand {\YNN}      {\mbox{$Y_{\text{NN}}$}}

\newcommand {\siaupm}   {\mbox{Si+Au$\rightarrow 2\pi^-$}}
\newcommand {\siaupp}   {\mbox{Si+Au$\rightarrow 2\pi^+$}}
\newcommand {\auaupm}   {\mbox{Au+Au$\rightarrow 2\pi^-$}}

\newcommand {\ANL}      {\mbox{Physics Division, Argonne National Laboratory, Argonne, Illinois 60439-4843}}
\newcommand {\BNL}      {\mbox{Brookhaven National Laboratory, Upton, New York 11973}}
\newcommand {\SSL}      {\mbox{Space Sciences Laboratory, University of California at Berkeley, Berkeley, California 94720}}
\newcommand {\CU}       {\mbox{Columbia University, New York, New York 10027 and Nevis Laboratories, Irvington, New York 10533}}
\newcommand {\UCR}      {\mbox{University of California at Riverside, Riverside, California 92507}}
\newcommand {\KEKtan}   {\mbox{High Energy Accelerator Research Organization (KEK), Tanashi-branch, Tokyo 188, Japan}}
\newcommand {\Kyushu}   {\mbox{Kyushu University, Fukuoka 812, Japan}}
\newcommand {\Kyoto}    {\mbox{Kyoto University, Sakyo-Ku, Kyoto 606, Japan}}
\newcommand {\LLNL}     {\mbox{Lawrence Livermore National Laboratory, Livermore, California 94550}}
\newcommand {\MIT}      {\mbox{Laboratory for Nuclear Science, Massachusetts Institute of Technology, Cambridge, Massachusetts 02139}}
\newcommand {\NYU}      {\mbox{New York University, New York, New York 10003}}
\newcommand {\Tokyo}    {\mbox{Department of Physics, University of Tokyo, Tokyo 113, Japan}}
\newcommand {\Tsukuba}  {\mbox{University of Tsukuba, Tsukuba, Ibaraki 305, Japan}}
\newcommand {\CNS}      {\mbox{Center for Nuclear Study, School of Science, University of Tokyo, Tanashi, Tokyo 188, Japan}}
\newcommand {\UM}       {\mbox{University of Maryland, College Park,
maryland 20742}}
\newcommand {\aLLNL}     {LLNL, Livermore, CA 94550}
\newcommand {\aBNL}      {BNL, Upton, NY 11973}
\newcommand {\aORNL}     {ORNL,, Oak Ridge, TN 37831}
\newcommand {\aRIKEN}    {RIKEN, Saitama 351-01, Japan}
\newcommand {\aYonsei}   {Yonsei Univ., Seoul 120-749, Korea}
\newcommand {\aKEK}      {KEK, 1-1 Oho, Tsukuba, Ibaraki 305 Japan}
\newcommand {\aISU}      {Iowa State Univ., Ames, IA 50010}
\newcommand {\aLBL}      {LBNL, Berkeley, CA 94720}
\newcommand {\aCERN}     {CERN, CH-1211, Geneve 23, Switzerland}
\newcommand {\aPurdue}   {Purdue Univ., West Lafayette, IN 47907}
\newcommand {\aDOE}      {U.S. Dept. of Energy, Germantown, MD 20874}
\newcommand {\aYale}     {Yale Univ., New Haven, CT 06520}
\newcommand {\aUIC}      {Univ. of Illinois, Chicago, IL 60607}
\newcommand {\aBergen}   {Univ. of Bergen, 5007 Bergen, Norway}
\newcommand {\aGSI}      {GSI, D-64291 Darmstadt, Germany}

\begin{document}

\date{\today} 
\title{System, centrality, and transverse mass dependence of two-pion
correlation radii in heavy ion collisions at 11.6 and 14.6 \AGeV}

\affiliation{\ANL}
\affiliation{\BNL}
\affiliation{\SSL}
\affiliation{\CU}
\affiliation{\UCR}
\affiliation{\KEKtan}
\affiliation{\Kyushu}
\affiliation{\Kyoto}
\affiliation{\LLNL}
\affiliation{\UM}
\affiliation{\MIT}
\affiliation{\NYU}
\affiliation{\Tokyo}
\affiliation{\Tsukuba}
\affiliation{\CNS}

\author{L.~Ahle}\altaffiliation[Present address:]{\aLLNL}\affiliation{\MIT}
\author{Y.~Akiba}\affiliation{\KEKtan} 
\author{K.~Ashktorab}\affiliation{\BNL} 
\author{M.{\,}D.~Baker}\altaffiliation[Present address:]{\aBNL}\affiliation{\MIT}
\author{D.~Beavis}\affiliation{\BNL} 
\author{P.~Beery}\affiliation{\UCR} 
\author{H.{\,}C.~Britt}\affiliation{\LLNL} 
\author{B.~Budick}\affiliation{\NYU}
\author{J.~Chang}\affiliation{\UCR} 
\author{C.~Chasman}\affiliation{\BNL} 
\author{Z.~Chen}\affiliation{\BNL} 
\author{C.{\,}Y.~Chi}\affiliation{\CU} 
\author{Y.{\,}Y.~Chu}\affiliation{\BNL} 
\author{V.~Cianciolo}\altaffiliation[Present address:]{\aORNL}\affiliation{\MIT}
\author{B.{\,}A.~Cole}\affiliation{\CU} 
\author{J.{\,}B.~Costales}\affiliation{\MIT} 
\author{H.{\,}J.~Crawford}\affiliation{\SSL} 
\author{J.{\,}B.~Cumming}\affiliation{\BNL} 
\author{R.~Debbe}\affiliation{\BNL} 
\author{J.{\,}C.~Dunlop}\altaffiliation[Present address:]{\aYale}\affiliation{\MIT} 
\author{W.~Eldredge}\affiliation{\UCR} 
\author{J.~Engelage}\affiliation{\SSL} 
\author{S.{\,}Y.~Fung}\affiliation{\UCR} 
\author{E.~Garcia}\altaffiliation[Present address:]{\aUIC}\affiliation{\UM}
\author{M.~Gonin}\affiliation{\BNL} 
\author{S.~Gushue}\affiliation{\BNL} 
\author{H.~Hamagaki}\affiliation{\CNS} 
\author{L.{\,}F.~Hansen}\affiliation{\LLNL} 
\author{R.{\,}S.~Hayano}\affiliation{\Tokyo} 
\author{S.~Hayashi}\affiliation{\BNL} 
\author{G.~Heintzelman}\affiliation{\MIT} 
\author{S.~Homma}\affiliation{\KEKtan} 
\author{E.~Judd}\affiliation{\MIT}
\author{H.~Kaneko}\affiliation{\Kyoto} 
\author{J.~Kang}\altaffiliation[Present address:]{\aYonsei}\affiliation{\UCR}
\author{S.~Kaufman}\affiliation{\ANL} 
\author{W.{\,}L.~Kehoe}\affiliation{\MIT} 
\author{E.{\,}J.~Kim}\affiliation{\BNL}
\author{A.~Kumagai}\affiliation{\Tsukuba} 
\author{K.~Kurita}\altaffiliation[Present address:]{\aRIKEN}\affiliation{\CU}
\author{R.{\,}J.~Ledoux}\affiliation{\MIT} 
\author{J.{\,}H.~Lee}\affiliation{\BNL} 
\author{M.{\,}J.~LeVine}\affiliation{\BNL} 
\author{J.~Luke}\affiliation{\LLNL} 
\author{Y.~Miake}\affiliation{\Tsukuba} 
\author{A.~Mignerey}\affiliation{\UM}
\author{D.{\,}P.~Morrison}\altaffiliation[Present address:]{\aBNL}\affiliation{\MIT}
\author{R.{\,}J.~Morse}\affiliation{\MIT} 
\author{B.~Moskowitz}\affiliation{\BNL} 
\author{M.~Moulson}\affiliation{\CU} 
\author{C.~M\"{u}ntz}\altaffiliation[Present address:]{\aGSI}\affiliation{\BNL}
\author{S.~Nagamiya}\altaffiliation[Present address:]{\aKEK}\affiliation{\CU}
\author{M.{\,}N.~Namboodiri}\affiliation{\LLNL} 
\author{T.{\,}K.~Nayak}\affiliation{\CU} 
\author{C.{\,}A.~Ogilvie}\altaffiliation[Present address:]{\aISU}\affiliation{\MIT}
\author{J.~Olness}\affiliation{\BNL} 
\author{C.{\,}G.~Parsons}\affiliation{\MIT} 
\author{L.{\,}P.~Remsberg}\affiliation{\BNL}
\author{D.~Roehrich}\altaffiliation[Present address:]{\aBergen}\affiliation{\BNL} 
\author{P.~Rothschild}\affiliation{\MIT} 
\author{H.~Sako}\affiliation{\CNS} 
\author{H.~Sakurai}\affiliation{\Tokyo} 
\author{T.{\,}C.~Sangster}\affiliation{\LLNL} 
\author{R.~Seto}\affiliation{\UCR} 
\author{J.~Shea}\affiliation{\UM}
\author{K.~Shigaki}\affiliation{\KEKtan}
\author{R.{\,}A.~Soltz}\altaffiliation[Present address:]{\aLLNL} \affiliation{\MIT}
\author{P.~Stankus}\altaffiliation[Present address:]{\aORNL}\affiliation{\CU}
\author{S.{\,}G.~Steadman}\altaffiliation[Present address:]{\aDOE}\affiliation{\MIT}
\author{G.{\,}S.{\,}F.~Stephans}\affiliation{\MIT} 
\author{T.{\,}W.~Sung}\affiliation{\MIT} 
\author{Y.~Tanaka}\affiliation{\Kyushu}
\author{M.{\,}J.~Tannenbaum}\affiliation{\BNL} 
\author{J.{\,}H.~Thomas}\altaffiliation[Present address:]{\aLBL}\affiliation{\LLNL}
\author{S.{\,}R.~Tonse}\altaffiliation[Present address:]{\aLBL}\affiliation{\LLNL}
\author{S.~Ueno-Hayashi}\affiliation{\Tsukuba} 
\author{J.{\,}H.~van Dijk}\affiliation{\BNL} 
\author{F.~Videb{\ae}k}\affiliation{\BNL} 
\author{O.~Vossnack}\altaffiliation[Present address:]{\aCERN}\affiliation{\CU}
\author{V.~Vutsadakis}\affiliation{\MIT} 
\author{F.~Wang}\altaffiliation[Present address:]{\aPurdue}\affiliation{\CU}
\author{Y.~Wang}\affiliation{\CU} 
\author{H.{\,}E.~Wegner}\affiliation{\BNL} 
\author{D.~Woodruff}\affiliation{\MIT} 
\author{Y.~Wu}\affiliation{\CU} 
\author{H.~Xiang}\affiliation{\UCR}
\author{G.{\,}H.~Xu}\affiliation{\UCR} 
\author{K.~Yagi}\affiliation{\Tsukuba} 
\author{X.~Yang}\affiliation{\CU} 
\author{H.~Yao}\affiliation{\MIT}
\author{D.~Zachary}\affiliation{\MIT} 
\author{W.{\,}A.~Zajc}\affiliation{\CU} 
\author{F.~Zhu}\affiliation{\BNL} 

\collaboration{E-802 Collaboration}

\begin{abstract}
Two-pion correlation functions are analyzed at mid-rapidity for three
systems (14.6~\AGeV~Si+Al, Si+Au, and 11.6~\AGeV~Au+Au), seven distinct
centrality conditions, and different \kT~bins in the range
0.1--0.5~GeV/c.  Source reference frames are determined from fits to
the Yano-Koonin source parameterization.  Bertsch-Pratt radius
parameters are shown to scale linearly with both number of 
projectile and total participants as obtained from a Glauber model
calculation.  A finite emission duration that increases linearly with
system/centrality is also reported.  The \mT~dependence of the
Bertsch-Pratt radii is measured for the central Si+Au and Au+Au
systems.  The system/centrality dependence is investigated separately
for both high and low \mT~regions.
\end{abstract}


\maketitle


\section{Introduction}
\label{sec:intro}

Bose-Einstein correlations of identical charged pions were first
observed and used to extract a source size for pion emission in
\ppbar~annihilations~\cite{gol60}.  This technique has since
been applied to heavy ion collisions ranging in energy from 1.8~\AGeV\
at the Bevalac~\cite{fun78,zaj84} to $\sqrt{s_{\text{NN}}}=130$~GeV at
RHIC~\cite{starprl01,phenixprl} (see also recent
reviews~\cite{cso02,wie99} and references therein).  Symmetrization of
the two-pion wave function results in an enhancement of the
two-particle correlation in a region of low relative momentum, the
extent of which is inversely proportional to the size or radius of the
emitting source.  The technique is most commonly referred to as HBT,
after the similar technique pioneered by Hanbury-Brown and Twiss to
measure stellar radii from intensity interferometry with radio
waves~\cite{han54}.  The form of the two-pion correlation depends upon
the emission function assumed for the pion source.  For static source
distributions with no final state interactions, such as the Fermi
Statistical Model first used by Goldhaber~{\it et al.}~\cite{gol60},
the correlation function is related to the square of the 3-dimensional
Fourier transform of the source distribution, $\rho ({\bf r})$, with
respect to the pion relative momentum, ${\bf q} = {\bf p_1} - {\bf
p_2}$,
\begin{equation}
C({\bf p_1},{\bf p_2}) \equiv
\frac{P({\bf p_1},{\bf p_2})}{P({\bf p_1})P({\bf p_2})} = 
1 + |\tilde{\rho}({\bf q})|^2.
\label{eq:FT}
\end{equation}
For a multi-dimensional, Gaussian parameterization of the source, the
enhancement in the relative momentum correlation is given by a
multi-dimensional Gaussian with each Gaussian width inversely
proportional to the canonically conjugate dimension, or radius.  For
dynamic, rapidly expanding sources the radii decrease with increasing
transverse mass, \mT, of the pion
pair~\cite{pra84,pra86,mak88,cha95ext,wie96}, where
$\mT=\sqrt{\kT^2+m^2_{\pi}}$, and \kT~is the mean transverse momentum
of the pion pair.  These \mT-dependent radii correspond to
the relative separations of the pions with low relative momentum and
are commonly referred to as ``lengths of homogeneity''.

Interest in measuring source sizes in heavy ion collisions comes
partly from the expectation that the presence of a QCD phase
transition will lead to large source sizes and long lifetimes for
particle emission~\cite{pra86,ber89,ris96}.  The absence of such a
signal, coupled with the complex and subtle nature of the \mT\
dependence, has left many puzzled over how best to interpret the
available HBT results.  However, in heavy ion collisions, there exists
additional geometric information from the size of the system, the
centrality of the collision, and also the reaction plane.  Results on
the pion source shape relative to the reaction plane for non-central
collisions~\cite{lis00,wie98} provide one important confirmation that
HBT radii are strongly correlated with physical dimensions of the
nuclear overlap region.  A previous comparison of HBT sources to nuclear
geometry~\cite{bar86} also confirmed this correlation, but the
compilation included different source parameterizations, online
triggers, and experimental acceptances.

We present a study of the dependence of the radii on system size and
centrality for the two-pion correlation function from
14.6~\AGeV~Si+Al, Si+Au, and 11.6~\AGeV~Au+Au collisions at the BNL
AGS measured by the E802 rotating magnetic spectrometer (Henry
Higgins).  The radii are compared to three geometric quantities: the
effective nuclear radius for the number of projectile participants,
the radius for the total participants, and the transverse distribution
of binary collisions calculated with a Glauber model.  The
data sets of the three systems span a range of seven distinct nuclear
geometries.  The \mT\ dependence is examined for the high statistics
central Si+Au and Au+Au systems and compared to measurements of Au+Au
central collisions by the E866 Forward Spectrometer.  Furthermore, the
centrality dependence is investigated separately for high and low
values of \mT\ for all systems.

Sec.~\ref{sec:exp} describes the experimental apparatus relevant to
this analysis.  Sec.~\ref{sec:redux} covers the data reduction.  The
correlation analysis procedure is described in Sec.~\ref{sec:analysis}
and results are presented in Sec.~\ref{sec:res}.  The conclusions are
given in Sec.~\ref{sec:conc}.  Appendix~\ref{sec:tables} contains the
complete set of tables for the correlation radii.  A Monte Carlo study
to check systematic effects in the Yano-Koonin fits is presented in
Appendix~\ref{sec:ykmc}.  Details and systematic studies of the
Coulomb correction are provided in Appendix~\ref{sec:coulomb}.

\section{Experiment}
\label{sec:exp}

The Si+Al/Au data were collected in 1991 and 1992 by BNL E859, and the
Au+Au data were collected in 1992 and 1994 by BNL E866. Both
experiments were extensions of BNL E802.  This series of experiments
ran in the B1 line of the Brookhaven AGS from 1986 through 1995.  The
complete experimental setups are described
elsewhere~\cite{e802nim,e866nim}.  The components essential
for this analysis are a Target Multiplicity Array (TMA), a Zero-degree
Calorimeter (ZCAL)~\cite{e802zcal}, a series of four drift chambers (T1--T4), two
trigger chambers (TR1,TR2), two multi-wire proportional chambers
(TRF1,TRF2) a time of flight wall (TOF), beam counters (BC), and a
level-2 trigger (LVL2).  

The TMA measured total charged particle multiplicity from resistive
tube pads configured in a wall and barrel array surrounding the
target.  A sum of the discriminated pad signals was used in the
level-1 trigger to enhance both central and peripheral data sets.  The
TMA was present only for the Si beam and was removed prior to the Au
beam commissioning in 1992.  The ZCAL consisted of alternating layers
of Pb and scintillator and was placed 11~m downstream from the target.
It was completely rebuilt and incorporated into the level-1 trigger
prior to the start of Au beam in 1992.  For the data sets presented
here the rms energy resolution ranged from $2.0 \sqrt{E}$ for the Si
beam to $3.2 \sqrt{E}$ for the Au beam.

Beam definitions and $t_0$ for TOF were provided by upstream beam
counters (BC) in anti-coincidence with a series of veto paddles.  The
interaction trigger condition was established by a bullseye (BE)
scintillator detector, placed just before the ZCAL.  For the Au~beam,
this was replaced by a \v{C}erenkov radiator.  The interaction trigger
was defined by $Z<26.5$ for the Si Beam, and $Z<73$ for the
Au beam.

The four drift chambers were placed two before and two after the
magnet.  T1, T2, T3, and T4, were positioned approximately 1, 1.5, 4,
and 4.5~m downstream from the target, respectively.  The drift
chambers provided a position resolution of 150~$\mu$m, and a minimum
two-track separation of 2~mm.  For this analysis, the
magnet was run at 0.4~Tesla, for an integrated field of 0.585~T-m.
The trigger chambers, TR1 and TR2, were located behind T3 and T4,
respectively.  For the 1994 running, two highly segmented multi-wire
chambers, TRF1 and TRF2 were added behind T1 and T2, respectively, to
improve track reconstruction in the higher multiplicity environment.
Resolutions for the TRF chambers were approximately 200~$\mu$m.  The TOF
was placed about 6~m downstream and particle flight times were
measured with an rms resolution of 120~ps.

The level-2 trigger (LVL2) was an integral part of the data taking and
was especially important for collecting two-pion events in peripheral
collisions.  It consisted of TR1, TR2, TOF, and a series of LeCroy
CAMAC modules used for fast readout, hit storage, memory lookup, and
logic~\cite{zaj92lvl2}. The trigger algorithm looped over all
TR1-TOF hit combinations and matched hits on TR2 for tracks projected
from the target.  Momentum and $1/\beta$ for valid combinations were
determined through lookup tables based on hit positions and
time-of-flight.  One or two independent regions in mass vs. momentum could be
specified in any logical combination to form the trigger condition.
All events for which the LVL2 did not produce a veto within 40~$\mu$sec
were accepted.  A set of events in which the LVL2 decision was
recorded but not used to veto were collected for each LVL2 condition
used in the experiment.  All data sets presented here made use of the
level-2 trigger, sometimes in combination with a level-1 centrality
trigger in the TMA or ZCAL.  LVL2 rejection factors ranged from 3--10
for Au+Au central to Si+Al peripheral events.  Trigger inefficiencies
were determined to be of order 1\% for these data, and were dominated
by the wire chamber inefficiencies.  No detectable trigger bias was
found for the two-pion data sets.

Nine data sets were taken covering seven different system and
centrality conditions.  These data sets are summarized in order of
increasing system size and centrality in Table~\ref{tab:data}.  The
online cut column refers to the multiplicity hardware centrality
definition for the Si beam data and the forward energy definition for
the Au beam data.  Sets 1 and 2 are for $\pi^+$ pairs from a
14.6~\AGeV~Si beam incident on a 1.63~g/cm$^2$ Al target (6\% beam
interaction length). 
Set 1 combines both peripheral TMA triggered data and minimum bias
data.  Set 2 includes only the minimum bias data.  Sets
3--6~are for pion pairs from a Si beam incident on a
0.944~g/cm$^2$ Au target (1\%).  Two different TMA level-1
thresholds were used to collect negative pion pairs for peripheral and
semi-peripheral events (sets 3 and 4).  The central TMA trigger was
used to collect pion pairs of both signs (sets 5 and 6).  Negative
pion pairs from 11.6~\AGeV~Au+Au collisions were taken using the same
Au target (1.5\% interaction length for Au beam) for central triggers
in 1992 (set 9), and for minimum bias events in 1994 (sets 7 and 8).
\begin{table}
\caption{Summary of correlation data sets.}
\label{tab:data}
\begin{ruledtabular}
\begin{tabular}{c|l|c|c|r}
    &        & Online     & Offline    &       \\
Set & System & Centrality & Centrality & Pairs \\
\hline
1. & Si+Al$\rightarrow 2\pi^+$+X & periph./m.b. & \ntma$<$45   & 83602 \\
2. & Si+Al$\rightarrow 2\pi^+$+X & min. bias    & \ntma$>$30   & 78713 \\
\hline
3. & Si+Au$\rightarrow 2\pi^-$+X & periph.      & \ntma$<$75   & 50713 \\
4. & Si+Au$\rightarrow 2\pi^-$+X & semi-periph. & \ntma$<$115  & 98468 \\
5. & Si+Au$\rightarrow 2\pi^-$+X & central      & \ntma$>$75   & 232296\\
6. & Si+Au$\rightarrow 2\pi^+$+X & central      & \ntma$>$75   & 76661 \\
\hline
7. & Au+Au$\rightarrow 2\pi^-$+X & min. bias    & \ezcal$>$550 & 77837 \\
8. & Au+Au$\rightarrow 2\pi^-$+X & min. bias    & \ezcal$<$550 & 88198 \\
9. & Au+Au$\rightarrow 2\pi^-$+X & central      & none         & 85573  \\
\end{tabular}
\end{ruledtabular}
\end{table}
The Au+Au minimum bias data set was divided at \ezcal$=550$ in order
to match the online centrality condition for data set 9.  All data
sets were subject to offline cuts to reject beam pile-up and to
improve centrality definition.  Table~\ref{tab:data} gives the final
pion pair statistics after all offline cuts.
 
\begin{figure}[htb]
\includegraphics[width=3.5in]{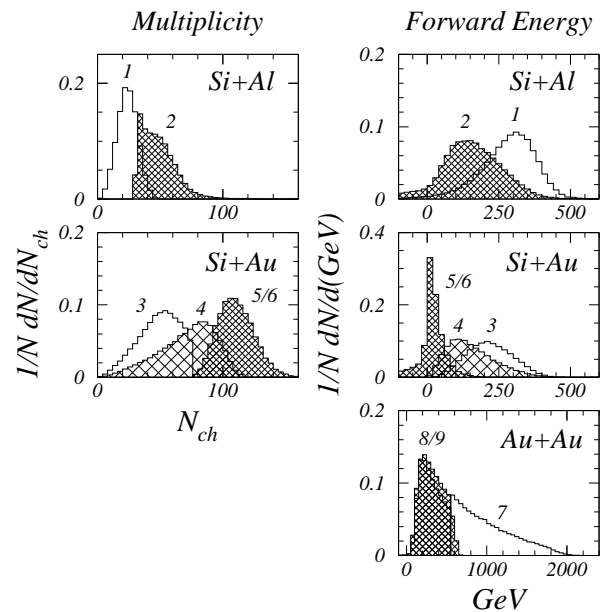}
 \caption{TMA multiplicity and ZCAL forward energy distributions for
 all pion pair events.  Central distributions are densely hatched,
 peripheral distributions are unfilled, and the Si+Au~semi-peripheral
 distribution is sparsely hatched.  Data set numbers from
 Table~\ref{tab:data} are printed alongside the distributions.  All
 distributions are normalized to unit area.}
\label{fig:multzcal}
\end{figure}

The final TMA multiplicity and ZCAL forward energy distributions for
all nine data sets are shown in Fig.~\ref{fig:multzcal}.  Only the
ZCAL distributions are shown for the Au+Au system (sets 7--9) --- a
new multiplicity array (NMA) was in place for the collection of data
in sets 7 and 8 but was not used in this analysis.  The negative
values in some of the \ezcal~distributions were the results of
excessive noise in the ZCAL electronics readout.  This effect is
taken into account in the centrality determination described in the
next section.

\section{Data Reduction}
\label{sec:redux}
 
\subsection{Centrality Determination}
\label{sec:cent}

The forward energy distributions from ZCAL were
used to determine the collision geometry for each system.  The number of
participant nucleons from the projectile is nominally calculated from
the ZCAL forward energy (\ezcal) and the kinetic energy per nucleon
(\enucl) according to,
\begin{equation}
  \ZCNproj = A - \frac{\ezcal}{\enucl}.
  \label{eq:Nproj}
\end{equation}
In applying Eq.~\ref{eq:Nproj} to the data, negative \ezcal~values
were included in the average.  The mean values for $\ZCNproj^{1/3}$
are given in Table~\ref{tab:geo}.

To account for the effect of the excess noise in the ZCAL, and to
obtain a more accurate determination of the collision geometry, a
Monte Carlo Glauber model was used to simulate the forward energy
distributions.  We use a value of 30~mb for the N-N cross-section and
the nuclear distributions are given by a Wood-Saxon distribution with
nuclear radius, $r=1.07 A^{1/3}$~fm, and surface thickness
0.55~fm~\cite{hah56}.

The ZCAL value for each Monte Carlo event in a given system was
calculated by sampling from a Poisson distribution set by the beam
distribution for the data.  This distribution was then smeared with a
Gaussian to simulate the additional noise observed in the
distributions.  The width of the Gaussian term (10--20~GeV) was chosen
to match the noise in the target-out corrected minimum bias
distribution while maintaining agreement for beam events.

\begin{figure}[htb]
\includegraphics[width=3.5in]{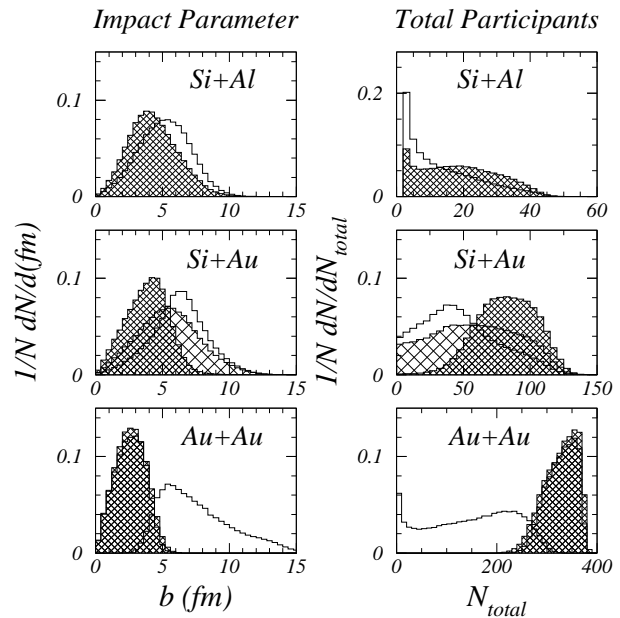}
 \caption{Distributions for impact parameter (left) and total
 participants (right) from Monte Carlo Glauber model for all data
 sets.  More central distributions are more densely hatched, as in
 Fig.~\ref{fig:multzcal}.}
\label{fig:nprojtot}
\end{figure}

The model was used to calculate distributions for the number of
projectile, target, and total participants and the transverse rms
distribution of binary collisions ($\rho_{\text{rms}}$).  Each Monte Carlo event was
assigned a weight according to it's calculated ZCAL value, defined by
the ZCAL distribution for 2$\pi$ events divided by the minimum bias
distribution.  This weighting procedure is necessary to reproduce the
correct ZCAL distribution for events with two pions in the spectrometer.
The weighted distributions for the impact parameter and
total participants are shown in Fig.~\ref{fig:nprojtot}.  The mean
values are listed in Table~\ref{tab:geo}.  The average value of
\ZCNproj~calculated with Eq.~\ref{eq:Nproj} are similar to the values
of \Nproj~calculated using the Glauber model.  All geometric
quantities are observed to increase with increasing centrality within
each system.  The geometric values are identical for the two central
Si+Au data sets and nearly equivalent for the two central Au+Au data
sets.
\begin{table}
\caption{Average geometric quantities
 calculated from ZCAL distributions and a Monte Carlo Glauber model.
 Participant numbers are raised to the one-third power and then
 averaged.  The value for $\rho_{\text{rms}}$ is the transverse rms
 distribution of all nucleon-nucleon interactions (in units of fm)
 averaged over all collisions.}
\label{tab:geo}
\begin{ruledtabular}
\begin{tabular}{c|c|c|c|c|c|c}
System & Set & \ZCRproj\ & \Rproj & \Rtarg & \Rpart & $\rho_{\text{rms}}$ \\
\hline
\multirow{2}*{Si+Al} 
 & 1 & 1.94 & 1.76  & 1.75  & 2.20  & 1.15  \\ 
 & 2 & 2.46 & 2.23  & 2.18  & 2.78  & 1.57  \\ 
\hline					    
\multirow{4}*{Si+Au} 			    
 & 3 & 2.19 & 2.10  & 2.43  & 2.87  & 1.59  \\ 
 & 4 & 2.49 & 2.43  & 2.94  & 3.42  & 1.85  \\ 
 & 5 & 2.87 & 2.88  & 3.79  & 4.29  & 2.20  \\ 
 & 6 & 2.87 & 2.88  & 3.79  & 4.29  & 2.21  \\ 
\hline					    
\multirow{3}*{Au+Au} 			    
 & 7 & 4.40 & 3.95  & 3.97  & 5.00  & 2.52  \\ 
 & 8 & 5.52 & 5.43  & 5.44  & 6.84  & 3.35  \\ 
 & 9 & 5.45 & 5.48  & 5.46  & 6.89  & 3.36  \\ 
\end{tabular}
\end{ruledtabular}
\end{table}

\subsection{Pion identification and pair selection}
\label{sec:pion}

The tracking algorithm for the 1991-1992 data (sets 1--6,
9)~\cite{rot94} began with a road-finder to connect TOF, TR1 and TR2
hits.  Additional hits collected on the downstream tracking chambers
were used to construct the downstream track segment.  This was
projected through the magnet using an effective dipole field
approximation assuming a target vertex and hist were collected on T1 and T2.
All hits associated with the track were fit with the vertex constraint
removed.  Tracks were required to pass goodness of fit cuts and to
project to within 2~cm of the nominal target position.  The single
particle momentum resolution was determined by GEANT simulation and
parameterized as $\sigma_p = ap/\beta \oplus b p^2$, with
$a=1.2\%$ and $b=0.6\%$~(GeV/c)$^{-1}$.  For the 1994 data (sets 7--8)
upstream track 
segments were found and fit independently using the additional hits on
TRF1 and TRF2.  The two track segments were then fit using the same
magnetic field approximation, and applying the same vertex and
goodness of fit cuts.  Using additional information from TRF1 and TRF2
the momentum resolution for 1994 was improved to give $a=0.85\%$ and
$b=0.12\%$~(GeV/c)$^{-1}$~\cite{wan96}.

\begin{figure}[htb]
\
\includegraphics[width=3in]{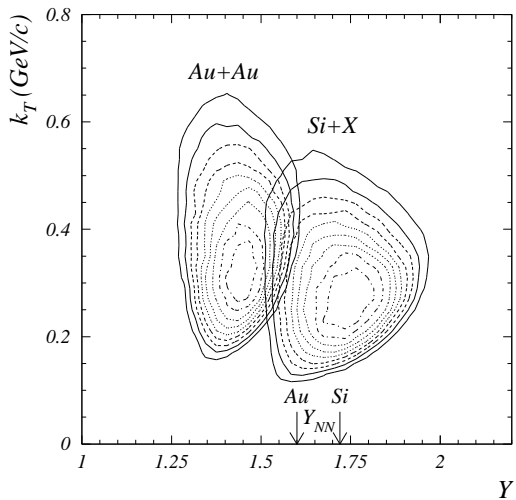}
 \caption{Distributions in \kT\ and rapidity for Si+X
 (14\dg~spectrometer setting) and Au+Au (21\dg~setting).  Arrows
 indicate rapidity for nucleon-nucleon center of mass for
 14.6~\AGeV~Si and 11.6~\AGeV~Au beams.}
\label{fig:ykt}
\end{figure}

Pions were identified up to a momentum of 1.85~GeV/c by requiring a
unique association with a TOF hit, and a measured $1/\beta$ that lies
within $3\sigma$ of the expected value for pions and outside of the
$3\sigma$ band for kaons.  The dominant background comes from electron
contamination, determined to be less than 5\%~\cite{sol94} in the
region $0.54<p<1.3$~GeV/c.  Pion pair momentum distributions for the
central Si+Au and Au+Au data sets are shown in Fig.~\ref{fig:ykt}.
The Si-beam data were taken with a spectrometer setting of 14~degrees,
which covers an angular region from 14 to 28~degrees and leads to a
mean pair-rapidity for pions approximately equal to \YNN$=$1.72.  For
the Au-beam the spectrometer was rotated to 21~degrees, primarily to
achieve lower track densities in the upstream chambers.  The mean pion
pair-rapidity for the Au-beam data is 1.43, slightly backwards 
of $\YNN=1.61$.

The two-track resolution was determined separately for the 1991--1992
and 1994 tracking algorithms, by examining the ratio of signal to
event-mixed pairs versus the relative slopes
$\delta\theta_x$ and $\delta\theta_y$ upstream of the magnet.
For the 1991--1992 data set, the two-track efficiency is well parameterized
by a two-dimensional Gaussian,
\begin{eqnarray}
  \label{eq:2treff92}
  \varepsilon(\delta\theta_x,\delta\theta_y) & =  & 1 - 0.86 \exp{\left[
        - \frac{(2.75)^2 \delta\theta_x^2 + \delta\theta_y^2}
               {2(0.0121)^2}\right]}, \nonumber \\
  \delta\theta_x & = & p_{1x}/p_{1z} - p_{2x}/p_{2z},  \nonumber \\
  \delta\theta_y & = & p_{1y}/p_{1z} - p_{2y}/p_{2z}.
\end{eqnarray}
A detailed GEANT simulation of two-pion events produced the same
Gaussian widths for the two-track efficiency, but the coefficient for
the exponential was closer to unity.  One expects this coefficient to
be less than unity in the data due to the presence of mis-associated
hits from other tracks.
The inverse of Eq.~\ref{eq:2treff92} was applied as a correction in 
the region where the efficiency was greater than $\sim$50\%, defined as,
\begin{equation}
  \label{eq:2trcut92}
  \sqrt{(2.75)^2 \delta\theta_x^2 + \delta\theta_y^2} > 0.011.
\end{equation}
Pairs not in this region were removed from the analysis.

For the 1994 data (sets 7 and 8), the two-track efficiency is flat but
then drops sharply for close tracks.  For these data, no correction
was applied and the two-track cut is given by,
\begin{equation}
  \label{eq:2trcut94}
  \sqrt{\delta\theta_x^2 + \delta\theta_y^2} > 0.008.
\end{equation}

\section{Correlation Analysis}
\label{sec:analysis}

\subsection{Parameterizations}

Ideally the Bose-Einstein correlation function is defined to be the
ratio of the probability for emitting two pions in the presence of
Bose-Einstein symmetrization to the probability for the case with no
symmetrization.  The latter can be approximated using Monte Carlo or
event mixing~\cite{kop74} techniques.  For each data set, an
event-mixed background was formed from pion pairs from different
events chosen at random from the entire data set.  The number of
background pairs was chosen to be 5--10 times the signal to achieve
statistical errors smaller than those for the signal, but still Poisson
distributed~\cite{sol94}.  This background differs from the ideal in
two respects, it contains a residual correlation distortion first
noted by Zajc~\cite{zaj84}, and it lacks two-body final state
interactions such as the Coulomb interaction.  The effect of the
residual correlation is included in the systematic error estimates.
We correct for the Coulomb interaction using an analytic approximation
for the Coulomb wave function of two-pions emanating from a finite static
source.  This source is assumed to be a spherically symmetric Gaussian
in the pair center-of-mass frame.  This approximation has been
described before~\cite{e802qm96} and is reviewed in
Appendix~\ref{sec:coulomb}.  The two-track cuts
(Eq.~\ref{eq:2trcut92}~and~\ref{eq:2trcut94}) are applied to both the
signal and background.  Other final state interactions such as the
di-pion strong interaction~\cite{suz87,bow88} and the Coulomb
interaction with the nuclear remnant are expected to be
negligible~\cite{kop72} and are not considered.  The correlation
function is then given by,
\begin{equation}
C_{2}({\bf q,\kT}) = \frac{A({\bf q,\kT})}{B({\bf q,\kT})}
         \cdot   \frac{1}{\varepsilon(\delta\theta_x,\delta\theta_y)}
         \cdot   \frac{1}{F(\eta,\Qinv,\Rinv)}.
\label{eq:c2}
\end{equation}
Here, $A$ and $B$ are the signal and event-mixed background,
$\varepsilon$ is the two-track efficiency defined by 
Eq.~\ref{eq:2treff92}, and $F$ is the Coulomb correction defined by
Eq.~\ref{eq:coulmdb}.  \Qinv~is the magnitude of two-pion relative
4-momentum in the pair rest frame, and \Rinv~is the parameter returned
by the 1-D invariant Gaussian parameterization,
\begin{equation}
C_{2}(\Qinv) = 1 + \lambda e^{-(Q_{\text{inv}}^2 R_{\text{inv}}^2)}.
\label{eq:qinv}
\end{equation}
Fits to this parameterization are used to perform an iterative
determination of the Coulomb correction.  Beginning with the
assumption of a point source (i.e. Gamow correction), the 
\Rinv~parameter converges to within 1\% after two iterations for all
data sets.  

The Yano-Koonin parameterization~\cite{yan78} is used to extract the
longitudinal velocity of the source,
\begin{equation}
C_{2}(\Qinv,\ql ) = 1 + \lambda 
e^{- \left( R^2 +\tau^2 \right) \gamma^2_s \left( q_0 - \beta_s q_{\text{l}} \right)^2
   + \left( Q^2_{\text{inv}} R^2 \right) },
\label{eq:yk}
\end{equation}
where $\ql = p_{z1} - p_{z2}$ and $\beta_s$ is the longitudinal
component of the source velocity.  $R$ and $\tau$ are the spherically
symmetric Gaussian radius and emission duration, respectively.

Using the source frames determined from the Yano-Koonin fits, all data
sets were fit to the 3-D Bertsch-Pratt parameterization
~\cite{pra86,ber89},
\begin{equation}
C_{2}(q_{\text{l}},q_{\text{s}},q_{\text{o}}) = 1 + \lambda e^{\left(
   - q_{\text{l}}^2 R_{\text{l}}^2 
   - q_{\text{s}}^2 R_{\text{s}}^2 
   - q_{\text{o}}^2 R_{\text{o}}^2 \right) },
\label{eq:bp}
\end{equation}
where \ql~is the longitudinal component of the relative momentum from
Eq.~\ref{eq:yk}.  The transverse relative momentum is separated into
components perpendicular (\qs) and parallel (\qo) to the pion pair
velocity.  This parameterization measures the longitudinal and
transverse correlation lengths, \Rl~and \Rs, respectively.  For an
azimuthally symmetric source without dynamical correlations, the pion
emission duration is calculated from \Ro, \Rs, and the transverse
velocity of the pion pair, $\beta_T$,
\begin{equation}
\tau = \sqrt{\Ro^2 - \Rs^2}/<\beta_T>.
\label{eq:bptau}
\end{equation}
A variation of this parameterization suggested by
Chapman~\cite{cha95cross}, includes a cross-term, \Rlo, which is
expected to be non-vanishing for sources that are asymmetric in $z$,
\begin{equation}
C_{2}(q_{\text{l}},q_{\text{s}},q_{\text{o}}) = 1 + \lambda e^{\left( 
   - q_{\text{l}}^2 R_{\text{l}}^2 
   - q_{\text{s}}^2 R_{\text{s}}^2
   - q_{\text{o}}^2 R_{\text{o}}^2 
   - 2 q_{\text{l}} q_{\text{o}} R_{\text{lo}}^2 \right) }.
\label{eq:xtrm}
\end{equation}
It is instructive to relate the cross-term to the Generalized
Yano-Koonin parameterization (GYK)~\cite{cha95ext}, using the
relation, $\qzero = \vec{q} \cdot \vec{\beta}_{pr}$.
\begin{eqnarray}
\lefteqn{C_2(q_{\text{l}},q_{\text{T}},q_{\text{0}}) = 1 + \lambda e^{\left(
  - q_{\text{l}}^2R_{\text{l}}^2
  - q_{\text{T}}^2R_{\text{T}}^2 
  - q_{\text{0}}^2\tau^2 \right)}} \\
 & = & 1 + \lambda e^{\left( 
  - q_{\text{l}}^2 (R_{\text{l}}^2 + \beta_l^2 \tau^2)  
  - q_{\text{T}}^2 (R_{\text{T}}^2  + \beta_{T}^2 \tau^2) 
  - 2q_{\text{l}}q_{\text{o}}\beta_l \beta_T \tau^2) \right)
}. \nonumber
\label{eq:cross}
\end{eqnarray}
This provides a simple explanation for the cross-term,
\Rlo~in the absence of dynamical correlations, and illustrates why the
cross-term vanishes for sources which are symmetric about the $z=0$
plane. 

\subsection{Minimization}

A new log-likelihood minimization function is used in this analysis.
It assumes Poisson probability distributions for both signal and
background, with means given by $\mu$ and $\nu$ respectively.  The
correlation function is defined as the ratio of these means by
imposing a delta-function constraint on the conditional probability
for $C$, given $A$ and $B$,
\begin{eqnarray}
\label{eq:poisson}
P(C|A,B) & = & \int d\mu d\nu \left( \frac{\mu^A e^{-\mu}}{A!} \right)
                         \left( \frac{\nu^B e^{-\nu}}{B!} \right)
                         \delta \left( C - \mu/\nu \right) \nonumber \\
         & = & \frac{C^A}{A!B!} 
               \frac{\left( A+B+1\right) !}{\left( C+1\right)^{A+B+2}}.
\end{eqnarray}
Taking the negative of twice the log and collecting leading order
terms in $A$ and $B$ leads to the minimization function used in this
analysis,
\begin{equation}
\chisq_{P} = -2 \left[ 
A \ln \left( \frac{ C \left( A+B \right)}{A \left( C+1 \right)}\right)
+ B \ln \left( \frac{A+B}{B \left( C+1 \right)} \right) \right].
\label{eq:poifit}
\end{equation}
Note that Eq.~\ref{eq:poifit} approaches a chi-squared
distribution, $\frac{(A-CB)^2}{\sigma^2_A+\sigma^2_{CB}}$, in the
limit of large $A$ and $B$.  We compare Eq.~\ref{eq:poifit} to another
log-likelihood minimization function commonly used~\cite{zaj91},
\begin{eqnarray}
\lefteqn{\chi^2_{PML} = -2 ( C_2B - A ) - } \\
 & & 2 (\sigma^2_A + \sigma^2_{C_2 B}) 
     \ln \left( \frac{( C_2B - A ) + (\sigma^2_A + \sigma^2_{C_2 B})}
                     {(\sigma^2_A + \sigma^2_{C_2 B})} \right). \nonumber
\label{eq:pml}
\end{eqnarray}
This form also assumes a Poisson distributed signal for $A$, but the
error terms for the background have been derived by working backwards
from the chi-squared limit.  All terms in Eq.~\ref{eq:poifit}
follow naturally from the assumption of Poisson distributions for $A$
and $B$ and the log-likelihood method.  For this reason, the new
minimization function may be more accurate for fits involving bins
with small counts.  However, both
Eqs.~\ref{eq:poifit}~and~\ref{eq:pml} were shown to give consistent 
results in fits to the data.  The minimization was performed using the
MINUIT package~\cite{minuit}.

\section{Results}
\label{sec:res}

For each of the nine data sets in Table~\ref{tab:data} the three
parameterizations, \Qinv, Yano-Koonin, and Bertsch-Pratt were
performed in order.  The \Qinv~fits were used to determine the Coulomb
correction.  The longitudinal velocity of the source frames were
obtained from the Yano-Koonin fits, and the HBT radii from the
Bertsch-Pratt fits were used in analysis of centrality, system, and
\kT~dependence.  A complete set of tables for all fitted radii are
given in Appendix~\ref{sec:tables}.

\subsection{\Qinv~Fit Results}
\label{sec:qinv}

Fit results for the final iteration \Qinv\ fits are summarized in
Table~\ref{tab:qinv}.  Tables~\ref{tab:qinv} through~\ref{tab:bphi} are
found in Appendix~\ref{sec:tables}.  These radii were taken as input
to the Coulomb correction calculation for all subsequent multi-dimensional
parameterizations for these data sets.  For each system, \Rinv\
increases with collision centrality.  Table~\ref{tab:qinv} also lists
the mean transverse mass,\mmT, for each of the nine data sets.  In
calculating this mean, it is common to restrict the average to only
those pairs with low relative momentum, which carry the information on
the source dimensions.  However, for systems that vary in size, the
region of low relative momentum will also vary inversely with the
source size.  To account for this effect, the average restricted to
the region $\Qinv < 2 / \Rinv$ using the step function $\Theta$,
\begin{equation}
\label{eq:lowQ}
\mmTs = \frac
 {\sum_{i=1}^N \sqrt{(k^2_T)_i + m^2_{\pi}}
   \ \ \Theta(2/\Rinv - (\Qinv)_i)}
 {\sum_{i=1}^N
   \Theta(2/\Rinv - (\Qinv)_i)}
\end{equation}
The quantity \mmTs~should be sufficiently general to be used over a
wide variation of system sizes and experimental acceptances.
Table~\ref{tab:qinv} shows that \mmTs~is never less than 84\% of
\mmT~for these data.
\begin{figure}[htb]
\includegraphics[width=3.5in]{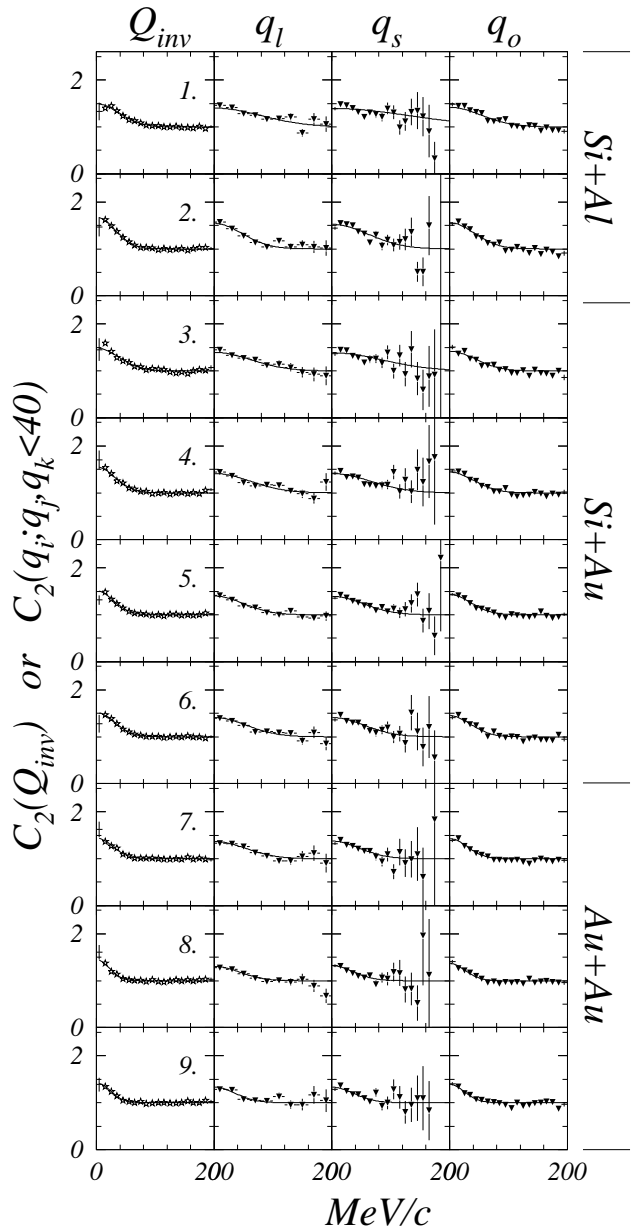}
 \caption{Correlation functions for data sets 1--9.  Data and
 parameterized fits for \Qinv~($\bigstar$) are shown in left panels.
 Right panels show slices of the correlation functions and fits for
 Bertsch-Pratt parameterization~($\blacktriangledown$) for
 $q<40$~MeV/c in the orthogonal variables.}
\label{fig:c2}
\end{figure}
The fully corrected \Qinv~correlation functions fits are shown in
the left set of panels of Fig.~\ref{fig:c2}.

\subsection{Yano-Koonin Fit Results}
\label{sec:yk}

By construction, the $R$ and $\tau$ parameters of Eq.~\ref{eq:yk}
are in the longitudinal rest frame of the source, and are invariant
under longitudinal boosts.  However, large longitudinal boosts yield
q-distributions that are extended along a narrow ridge of the
$\ql=\qzero$ axis, and this presents a challenge for binning and
minimization.  This effect was studied with a Monte Carlo simulation and
it was determined that there is a slight systematic bias in the fitted
$\tau$ and $\beta_s$ when the reference frame differs from the true
source frame by more than a half unit of rapidity.  The results of
this study are given in Appendix~\ref{sec:ykmc}.

To minimize this bias, all data sets were initially fit in the 
\YNN~frame, 1.72 for the Si-beam and 1.60 for Au.  Data sets with
values of \bsrc~which differed significantly from zero were then refit
in the frames expected to yield $\bsrc=0$.  Only the Si+Au central
data (sets 5 and 6) and Au+Au data (sets 7--9) required a second fit.
The weighted means for \bsrc for the Si+Au central and Au+Au systems
were $-0.24 \pm 0.03$ and $-0.15 \pm 0.05$ respectively.  This led to
the choice of the $Y=1.50$ frame from for central Si+Au, and the
$Y=1.45$ frame for Au+Au.  Final fit frames and all parameters for the
Yano-Koonin fits are listed in Table~\ref{tab:fityk}.  These frames
are used for all remaining fits.

All final values for \bsrc~are within two standard deviations of
zero. As with the \Qinv~results, the Yano-Koonin radius increases
monotonically with centrality for all systems and from smaller systems
to larger.  The emission duration shows a similar but weaker
dependence on system size and centrality.  We note that because the
acceptance for the Si-beam pairs centered at \YNN~(see
Fig.~\ref{fig:ykt}), the final values for \bsrc~for the Si+Al and
non-central Si+Au systems are consistent with both the assumptions of
a fixed source at \YNN~and longitudinal boost invariance.  The final
values of \bsrc~for the symmetric Au+Au system are closer to the mean
pair rapidity of $Y=1.43$ than to \YNN~and therefore imply a source
that is longitudinally boost invariant.  However, the values of
\bsrc~for the asymmetric Si+Au central system are indicative of a
pion source shifted backwards to $Y=1.50$ and are inconsistent with
the assumption of longitudinal boost-invariance.

\subsection{Bertsch-Pratt Fit Results}
\label{sec:bp}

The results of the Bertsch-Pratt fits with and without the Chapman
\Rlo\ cross-term are listed in Tables~\ref{tab:bp}~and~\ref{tab:rlo},
respectively.  The system and centrality dependence of the radii will
be presented graphically in Sec.~\ref{sec:centsys}.  The \Rlo\
cross-terms are consistent with zero for the symmetric systems
measured at mid-rapidity, as expected.  The largest values are for the
Si+Au and Au+Au central systems, and only for the central
\siaupm~does the \Rlo\ term differ from zero by more than $2\sigma$.

Slices of the correlation functions for the Bertsch-Pratt fits (no
cross-term) for all systems in Table~\ref{tab:bp} are shown in the
right three panels of Fig.~\ref{fig:c2}.  The slices in each
q-component are averaged over $q_i<40$~MeV/c in the orthogonal
q-variables.

\begin{figure}[htb]
\includegraphics[width=3in]{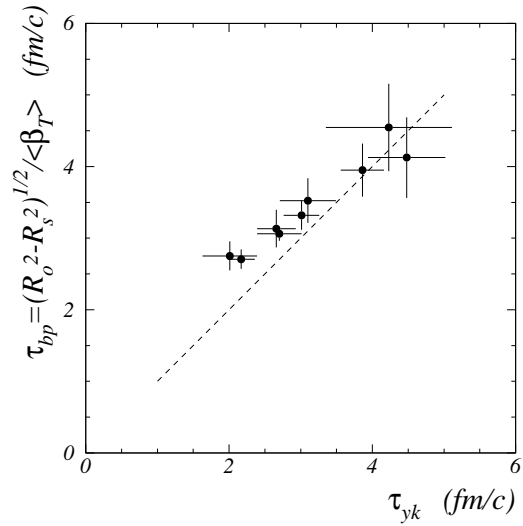}
 \caption{Comparison between $\tau$ determined from Yano-Koonin and
Bertsch-Pratt parameterizations.  The dashed line indicates
$\tau_{bp}=\tau_{yk}$.}
\label{fig:tvst}
\end{figure}

From the values of \Ro~and \Rs~in the Bertsch-Pratt fits we have
extracted values for the emission duration, $\tau$, using
Eq.~\ref{eq:bptau}.  These are compared to the corresponding emission
duration parameters measured in the Yano-Koonin fits.
Fig.~\ref{fig:tvst} shows the two parameterizations give similar
results, but with Yano-Koonin $\tau$ is smaller for the lighter
systems.  This trend is consistent with the assumption of spherical
symmetry in Yano-Koonin, and the trend towards greater asymmetry
(\Rl$>$\Rs) in the lighter systems.  The Generalized Yano-Koonin (GYK)
suggested in~\cite{cha95ext}, which separates the transverse and
longitudinal components of the source, is a logical extension to
Eq.~\ref{eq:yk}.  Use of this parameterization in the future may lead
to an improved correspondence between emission durations determined by
the two methods.  For the remaining sections $\tau$ is extracted from
Eq.~\ref{eq:bptau} and its error incorporates the \Ro-\Rs~covariant
term from the error matrix.

As a systematic check on the reference frames, Bertsch-Pratt fits were
also performed in the Longitudinal Co-Moving System (LCMS) frame,
defined as the local frame in which the longitudinal pair momentum is
zero.  These results are given in Table~\ref{tab:bplcms}.  All radii
are consistent with the fixed frame fits to within two standard
deviations.  Large differences are not expected given that the fixed
frames used in the previous fits were consistent with the pair LCMS
for all but the Si+Au central system, and the rms width of pair
rapidity distribution is small ($\sim$0.1).

\subsection{Centrality and System Dependence}
\label{sec:centsys}

\begin{figure}[htb]
\includegraphics[width=3.5in]{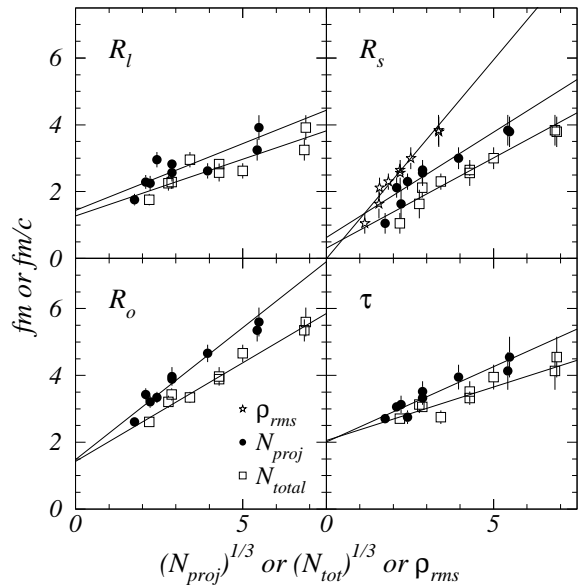}
 \caption{Bertsch-Pratt radii and emission duration
 vs. projectile participants (\Rproj), total participants (\Rpart),
 and the transverse rms distribution of binary collisions
 ($\rho_{\text{rms}}$).  All radii are fit to a linear dependence of
 the form $a+bx$, with fit values given in Table~III.}
\label{fig:scale}
\end{figure}

We examine the dependence of the HBT radii on system size and
centrality by comparing the Bertsch-Pratt radii given in
Table~\ref{tab:bp} to the geometric quantities of Table~\ref{tab:geo}.
Fig.~\ref{fig:scale} shows each of the radius parameters and the extracted
emission duration plotted vs. mean values of \Rproj~and \Rpart.  The
geometric dependences were fit to a linear form, $a+bx$.  The
extracted slopes and intercepts are given in Table~III.
\Rs~is also fit to to a linear dependence on $\rho_{\text{rms}}$,
yielding $a=-0.01 \pm 0.38$, $b= 1.19 \pm 0.18$, with a \chisq/dof of
3.5/7.  
\begin{table}
\label{tab:sc}
\caption{Linear fits of Bertsch-Pratt radii to \Nproj~and \Npart~determined from Glauber model calculations.}
\begin{ruledtabular}
\begin{tabular}{lr|c|c|c|c}
\multicolumn{2}{c|}{Fit} & \Rl & \Rs & \Ro & $\tau$ \\
\hline
\multirow{3}*{\Nproj}
& a & 1.44$\pm$0.21 & 0.63$\pm$0.31 & 1.48$\pm$0.19 & 2.01$\pm$0.23\\
& b & 0.40$\pm$0.07 & 0.63$\pm$0.10 & 0.79$\pm$0.07 & 0.45$\pm$0.10\\
& \chisq/dof & 18.2/7 & 8.4/7 & 10.5/7 & 5.6/7\\
\hline \hline
\multirow{3}*{\Npart}
& a & 1.27$\pm$0.22  & 0.31$\pm$0.34  & 1.43$\pm$0.19  & 2.05$\pm$0.22\\ 
& b & 0.34$\pm$0.05  & 0.54$\pm$0.08  & 0.59$\pm$0.05  & 0.32$\pm$0.07\\ 
& \chisq/dof & 12.6/7  & 2.8/7  & 6.3/7  & 6.3/7\\ 
\end{tabular}
\end{ruledtabular}
\end{table}
All fitted parameters exhibit a linear dependence on both \Rproj~and
\Rpart.  Converting \Rs~values to 3D rms radii~($\sqrt{3}\times \Rs$)
yields slopes of $0.99 \pm 0.14$. and $1.20 
\pm 0.17$ for \Rpart~and \Rproj, respectively.  The latter slope is
consistent with the $A_p^{1/3}$ scaling noted by
J.~Bartke~\cite{bar86}, where $A_p$ denotes the atomic number of the
projectile.  This scaling dependence was taken from an analysis of
mean-free paths of nuclear projectiles in emulsion
~\cite{tan85,hec78}.  It is equivalent to the A-dependence of a hard
sphere nuclear radius, and is consistent with measurements of the
nuclear charge distribution in electron scattering~\cite{hah56}.  The
\Rpart~slopes for \Rl~are 60\% smaller than for \Rs, whereas for \Ro~they
are 10\% larger.  In addition, the $\tau$ parameter also displays a
significant slope that is consistent with the dependence of \Rl.

We note that the rms values in reference~\cite{bar86} were derived from a
variety of trigger conditions, parameterizations and regions of \mT,
whereas in this analysis the parameterizations are uniform, the
trigger conditions are accounted for, and the value of \mmTs~is in the
range of 300--325~MeV for Si+Al/Au and 350--360~MeV for Au+Au.  The
implication of the higher \mmTs~range for Au+Au will be discussed in
the next section.

The \Rpart~fits have a slightly lower \chisq/dof than the
\Rproj~fits.  The success of the \Rpart, \Rproj~and
$\rho_{\text{rms}}$ fits illustrates that there is more than one way
to map the system/centrality onto geometry, and that the scaling of
HBT radii need not be restricted to an effective radius of an assembly of
wounded nucleons.

To translate the slopes and intercepts into a true geometric
dependence of the pion source sizes requires a careful study of the
\mT~dependence of each radius.  However, if the \mT~dependence is
largely independent of system/centrality, one can extrapolate the
entire scaling to lower \mT.  Then the scaling observed at moderate
values of \mT~will also be valid at lower \mT~value, where (for \Rs)
the geometric source is better approximated.

The intercepts for all \Rs~fits are consistent with zero, as one would
expect for the dependence of a transverse overlap region that is
extrapolated to zero participants.  The intercepts for all other
radius parameters differ significantly from zero.  This is physically
meaningful in the context of the standard interpretations given to
these parameters, where \Rl~is indicative of the hadronization time,
and \Ro~and $\tau$ are sensitive to the emission duration.  Thus, even
when extrapolating to zero participants, one may expect to measure
finite lengths and durations for the pion hadronization.

\subsection{Transverse Momentum Dependence}
\label{sec:mom}

\begin{figure}[htb]
\includegraphics[width=3.5in]{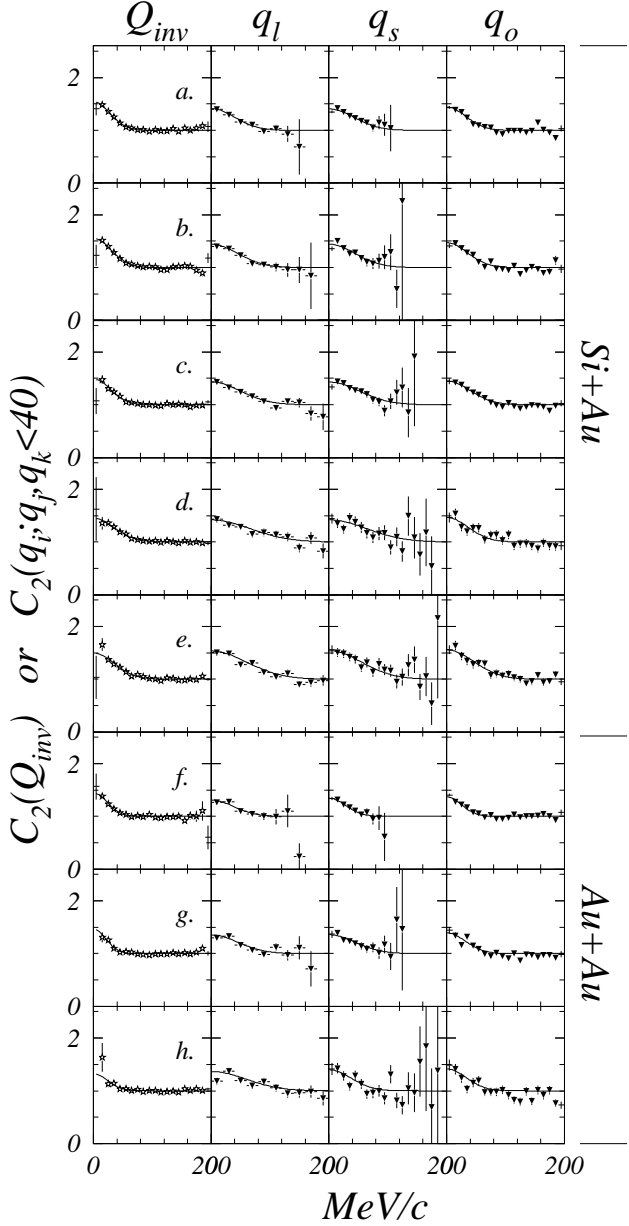}
 \caption{Correlation functions for \mT~data sets a--h.  Data and
 parameterized fits for \Qinv~($\bigstar$) are shown in left panels.
 Right panels show slices of the correlation functions and fits for
 Bertsch-Pratt parameterization~($\blacktriangledown$) for $q<40$~MeV/c 
 in the orthogonal variables.}
\label{fig:c2pt}
\end{figure}

To study the transverse momentum dependence the central \siaupm\ data
set was subdivided into three equal-statistics bins in \mT~(sets
a,c,e), and the \siaupp\ data set was subdivided into two equal bins
(sets b,d).  The two central \auaupm\ data sets were merged, and then
divided into three equal bins (sets f,g,h).  For this merged data set,
the two-track cut of Eq.~\ref{eq:2trcut94} was increased from 8~mrad
to 15~mrad, and no two-track correction was applied.  Fits to the full
merged data set were performed for all parameterizations above and
were found to be consistent with fits to the separate data sets.  The
data sets are given in Table~\ref{tab:bpkt}, along with the fit
parameters and values for \mmTs.  The fully corrected \Qinv~and
Bertsch-Pratt correlation functions and fits are shown in
Fig.~\ref{fig:c2pt}.

\begin{figure}[htb]
\includegraphics[width=3.5in]{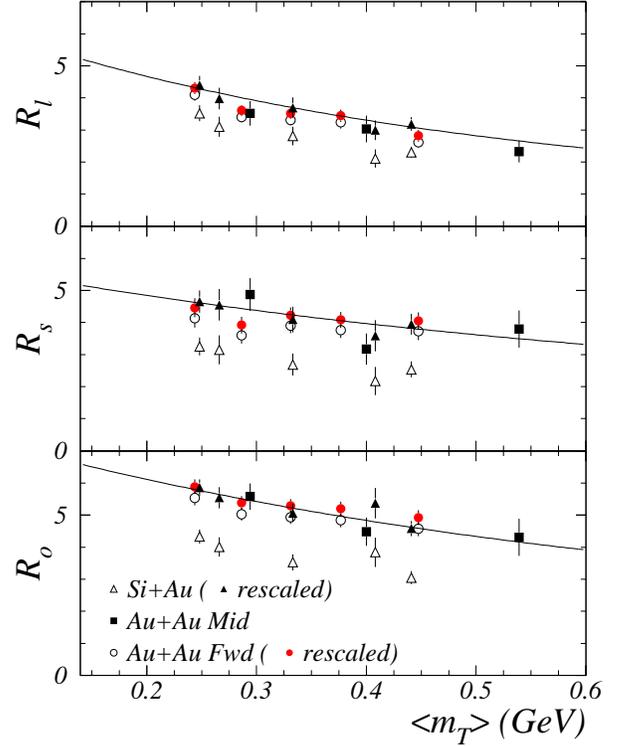}
 \caption{\mT~dependence of Bertsch-Pratt radii for Si+Au, Au+Au,
Au+Au at forward rapidity~\cite{jh98}, and for Si+Au rescaled by \Rpart~to match
Au+Au.  The solid line is a fit of the form $e^{(a+b m_T)}$ to the
Si+Au points, rescaled according the \Rpart~for Au+Au.  The Au+Au
forward rapidity data are also shown after being rescaled to match the
mid-rapidity centrality condition.}
\label{fig:ktnew}
\end{figure}

The transverse momentum dependence of the two-pion correlation
function has been discussed by many authors,
\cite{pra84,sin89,cha95mod,cha96rea,her95,wie96,wie97}, and in
principle can be used to test dynamic models for rapidly expanding
sources.  We have found that the small number of \kT~bins are
insufficient to constrain these models.  Instead we use an exponential
form, $e^{(\alpha+\beta m_T)}$ to provide an empirical description of the
\mT~dependence of the Si+Au system, and use this to check the Au+Au
system for consistency.  Values for $\alpha$ and $\beta$ are given in
Table~IV.  
Fig.~\ref{fig:ktnew} shows the \mT~dependence
for sets a--h, and for the Si+Au radii rescaled from $\Rpart = 2.87$
to $\Rpart=6.89$ to match Au+Au.  The solid line shows the
\mT~exponential fit also rescaled in this way.  The Au+Au data points are
consistent with the rescaled dependence for each radius.

Also shown are the E866 forward spectrometer results for $\pi^-$~pairs
from $1.9<Y<2.3$ for a centrality range that corresponds to the top
24\% of the interaction cross-section gated on
multiplicity~\cite{e866qm99,jh98}.  These data are less central than
the Au+Au data set at mid-rapidity, which correspond to the top
$\sim$15\% of the interaction cross-section gated on forward energy.
For comparison, the forward spectrometer results are also shown
rescaled up from $\Rpart=6.3$, as determined from the Glauber model
calculations for the 24\% most central collisions.  Unlike the
determination of $\Rpart$ for the other data sets, this was not done
using the forward energy distributions for events containing 2-pions.
This effect is largest for less central distributions, and therefore
the rescaling of the forward rapidity radii in Fig.~\ref{fig:ktnew}
should be taken as an upper bound.  Overall, there is good agreement
between the the rescaled Si+Au parameterized
\mT~dependence and the \mT~dependence of the Au+Au radii at both
forward and mid-rapidity.
\begin{table}
\label{tab:mtexp}
\caption{Exponential fit parameters for \mT\ dependence in central Si+Au.} 
\begin{ruledtabular}
\begin{tabular}{l|c|c|c}
Variable &  $\alpha$      &   $\beta$      &  \chisq/dof \\ 
\hline	    
\Rl      &  1.78$\pm$0.21 & -2.24$\pm$0.61 & 1.7/3  \\
\Rs      &  1.53$\pm$0.27 & -1.47$\pm$0.80 & 1.1/3  \\
\Ro      &  1.85$\pm$0.16 & -1.64$\pm$0.49 & 2.3/3  \\
$\tau$   &  1.59$\pm$0.64 & -1.76$\pm$2.05 & 4.9/3  \\
\end{tabular}
\end{ruledtabular}
\end{table}

The parameterized \mmTs~dependence of Si+Au radii can now be used to
check for errors arising from the 20\% higher \mmTs~values of the pion
pairs for Au+Au used in the centrality analysis.  The
\mmTs~dependence is used to interpolate all Bertsch-Pratt radii in
Table~\ref{tab:bp} to a common value of $\mmTs=0.325$~GeV and the
linear dependence on \Rpart~and \Rproj~is refit.  The results are
given in Table~V.  The slopes for all parameters
except $\tau$ increase slightly.  All difference are less than the
reported uncertainties, and the observations of the preceding section
with respect to slopes and intercepts remain valid.
\begin{table}
\label{tab:rescale}
\caption{Linear fits of Bertsch-Pratt radii to system/centrality
dependence interpolated to
$\mmTs=0.325$~GeV.}
\begin{ruledtabular}
\begin{tabular}{lr|c|c|c|c}
\multicolumn{2}{c|}{Fit} & \Rl & \Rs & \Ro & $\tau$ \\
\hline
\multirow{3}*{\Nproj}
& a & 1.23$\pm$0.21  & 0.45$\pm$0.31  & 1.29$\pm$0.20  & 1.91$\pm$0.23\\ 
& b & 0.47$\pm$0.07  & 0.69$\pm$0.10  & 0.86$\pm$0.07  & 0.49$\pm$0.10\\ 
& \chisq/dof & 11.9/7  & 7.0/7  & 8.4/7  & 7.7/7\\ 
\hline \hline
\multirow{3}*{\Npart}
& a & 1.11$\pm$0.22  & 0.16$\pm$0.34  & 1.31$\pm$0.20  & 2.01$\pm$0.23\\ 
& b & 0.38$\pm$0.06  & 0.57$\pm$0.08  & 0.62$\pm$0.05  & 0.33$\pm$0.07\\ 
& \chisq/dof & 9.1/7  & 2.1/7  & 11.9/7  & 10.5/7\\ 
\end{tabular}
\end{ruledtabular}
\end{table}

\subsection{Centrality and Momentum}
\label{sec:centmom}

\begin{figure}[htb]
\includegraphics[width=3.5in]{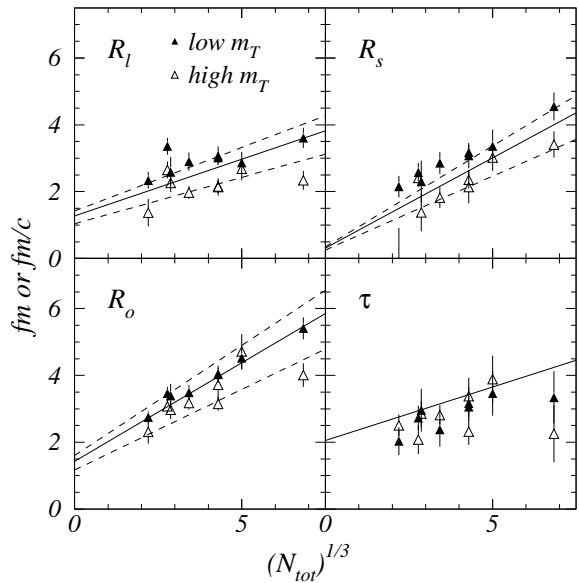}
 \caption{Bertsch-Pratt radii and emission extracted duration
 vs. total participants (\Rpart) for high \mT~($\triangle$) and low
 \mT~($\blacktriangle$)~bins.  The solid lines are linear fits to the
 full \mT~range, and the dashed lines are the same linear fits
 rescaled to \mmTs~of 275~MeV/c and 415~MeV/c.}
\label{fig:hilo}
\end{figure}

To investigate the centrality dependence for low and high regions of
\mT, each of the data sets 1--7, and 8+9 combined `was divided into
two \mT~bins of equal statistics.
The \mmTs, \Rinv, and Bertsch-Pratt radii for these fits are listed in
Tables~\ref{tab:bplo} and~\ref{tab:bphi}.  The system and centrality
dependence of the high and low \mT~regions are shown in
Fig.~\ref{fig:hilo}, plotted only for the \Npart~scaling.  Independent
linear fits to the two \mT~regions are unable to resolve significant
differences in the slopes and intercepts for the two regions.  The
data are compared to the \Rpart~linear fits (solid lines) 
for the full data sets that were shown in
Fig.~\ref{fig:scale}.  The dashed lines are equal to the solid line
fits interpolated up to 415~MeV and down to 275~MeV using the
parameterized \mT~dependence of Si+Au.

The high (low) \mT~values for \Rl~and \Rs~fall below (above) the
linear fits to full \mT~regions, and are approximately consistent with
the interpolations to higher (lower) \mT.  For \Ro~the high \mT~points
are closer to the solid line, and the low \mT~points fall slightly
lower.  For the $\tau$ parameter, nearly all points fall below the
unbinned scaling.  One possible explanation is a time-ordered pion
emission, in which the high \mT~pions are emitted at a time that is
displaced relative to the emission of the low
\mT~pions.  However, the potentially complex interplay between
space, time, and momentum must be considered fully.  At the very
least, these results suggest a direction for future theoretical work
and analysis with higher statistics measurements.

\begin{figure}[htb]
\includegraphics[width=3.5in]{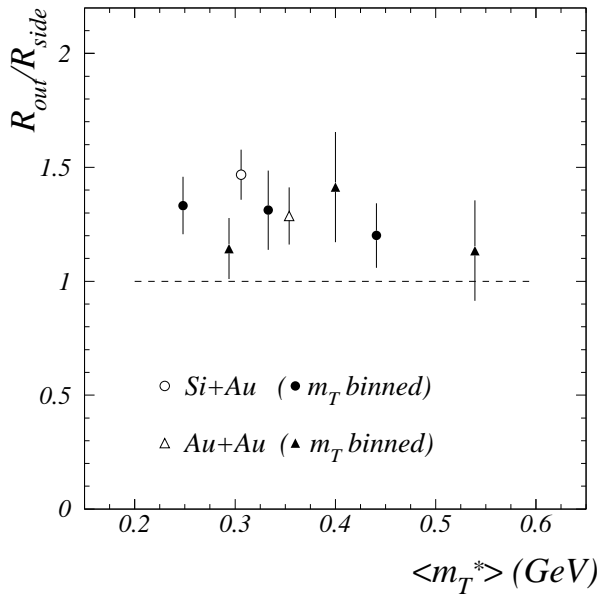}
 \caption{The ratio of \Ro/\Rs~for the high statistics data sets:
 central Si+Au (set 5, circles) and Au+Au (sets 8+9, triangles).  The
 open symbols are for the full \mT~range.  The filled symbols are for
 the subdivision into three bins for Si+Au (sets a, c, e) and Au+Au
 (sets g, f, h).}
\label{fig:outside}
\end{figure}

The ratio \Ro/\Rs~has been suggested as way to search for changes in
the pion emission duration~\cite{ris96} in heavy ion collisions.
This ratio is plotted in Fig.~\ref{fig:outside} for the unbinned 
Si+Au central 2$\pi^+$ (set 5) and Au+Au central data (set 8+9),
and for these data sets subdivided into three \mT~bins.
The values of \Ro/Rs~vary from 1.2 to 1.5, but do not indicate any
clear dependence on the \mT~bin size.   It may be interesting to
repeat this study in the future with larger data samples and`
statistically independent bins of different sizes.

\subsection{Systematic Errors}

The systematic error in determining the centrality is expected to be
dominated by the ZCAL noise.  In the course of modifying the noise
components in our model, we found nominal variations in the derived
mean quantities to be consistent with a systematic error of
approximately 5\%.

The sources of systematic errors for the HBT radii are summarized in
Table~\ref{tab:sys}, and nominal values are given.  The error due to
finite momentum resolution was examined for the \Rinv~fit of the Au+Au
central data set~\cite{kla97}, and the effect of residual correlations
in the background is reported elsewhere for the \Rinv~fit to the
central Si+Au data set~\cite{cia94,aki93}.  The variation for the
resolution smearing applies equally to each of the Bertsch-Pratt
radii.  All other variations are from \Rinv~parameterizations of the
central~\siaupm~data set.  The sign of the variation indicates the
direction of the effect of incorporating the correction/change.
\begin{table}
\caption{Sources and nominal values of systematic errors for the HBT
radii and $\lambda$.}
\label{tab:sys}
\begin{ruledtabular}
\begin{tabular}{l|d|d}
Source               & R      & \lambda\\
\hline
momentum resolution  &    +2\%  &    +4\% \\
5~MeV bins           & \pm 1\%  & \pm 2\% \\
residual correlation &    -2\%  &    +2\% \\
two-track correction & \pm 1\%  & \pm 2\% \\
Coulomb iteration    &    +0\%  &    -1\% \\
Coulomb resolution   &    -1\%  &    -1\% \\
\end{tabular}
\end{ruledtabular}
\end{table}
Although some of the systematic errors enter with different signs, the
different contributions are added in quadrature to yield a total
systematic error of 3\% for the radii and 5\% for $\lambda$.

The effect of a partial Coulomb correction is considered separately in
Appendix~\ref{sec:coulomb}.  If a significant fraction of pairs ($>50\%$)
are not subject to the Coulomb interaction, then this error would
dominate for both $R$ and $\lambda$.

\section{Conclusion}
\label{sec:conc}

These data demonstrate unambiguously the strong connection that exists
between the space-time measurements from Bose-Einstein correlations
and collision geometry.  All radii and lifetimes were observed to
depend significantly on several measurements of system size and
centrality.  At $\mmTs=0.325$~GeV, \Rs~was observed to scale with the
effective rms radius for the total number participants, as well
as the linear transverse size in a Glauber model.  For the \Rl~and
\myt~parameters the linear dependence on \Rpart~is 60\% weaker, and for
\Ro~it is 10\% stronger.  Only \Rs~extrapolates to an intercept of
zero.  The other parameters have finite intercepts of 1--2~fm for
$\Rproj,\Rpart=0.$ 

Yano-Koonin fits were used to determine the longitudinal source
velocities (\bsrc) for all systems.  For all symmetric and non-central
Si+Au systems measured at \YNN, \bsrc is consistent with both a fixed
source at \YNN~and a source which exhibits longitudinal boost
invariance.  The \bsrc~for the Au+Au systems are only consistent with
boost invariance, and for Si+Au central systems they imply a source
that is shifted backwards of \YNN.

The \mT~dependence of the radii has been parameterized for central
Si+Au and shown to be consistent with the central Au+Au \mT~dependence
at both mid and forward rapidity.  For the first time, the
system/centrality dependence was investigated independently for both
high and low \mT~regions.

The dependence of \Rl~and \Rs~with \Npart~at high and low \mT~is
consistent with interpolations based on the parameterized
\mT~dependence of the Si+Au central data.  However, the \Ro~and $\tau$
radii are inconsistent with this interpolated \mT~dependence, implying
that these two parameters are sensitive to the width of the \mT~bin.
Finally, the ratio \Ro/\Rs~is shown to be above unity for both the
Si+Au and Au+Au central systems, and binning the data in \mT~does not
change this result.

\begin{acknowledgements}
The work was supported by the U.S. Department of Energy under
contracts with ANL (W-31-109-ENG-38), BNL (DE-AC02-98CH10886),
Columbia University (DE-FG0286-ER40281), LLNL (W-7405-ENG-48), MIT
(DE-AC02-76ER03069), UC Riverside (DE-FG03-86ER40271), by NASA
(NGR-05003-513), under contract with the University of California, by
the Ministry of Education and KOSEF (951-0202-032-2), in Korea, and by
the Ministry of Education, Science, Sports, and Culture of Japan.  

We wish to thank E.P.~Hartouni and S.~Pratt for their support and for
enlightening discussions.  This paper is dedicated to the memory of
Chuck Parsons, whose insight, ingenuity, and enormously creative
spirit inspired a generation of graduate students on E859.
\end{acknowledgements}

\appendix

\section{Frame Dependence of the Yano-Koonin Parameterization}
\label{sec:ykmc}

\begin{figure}[htb]
\includegraphics[width=3.5in]{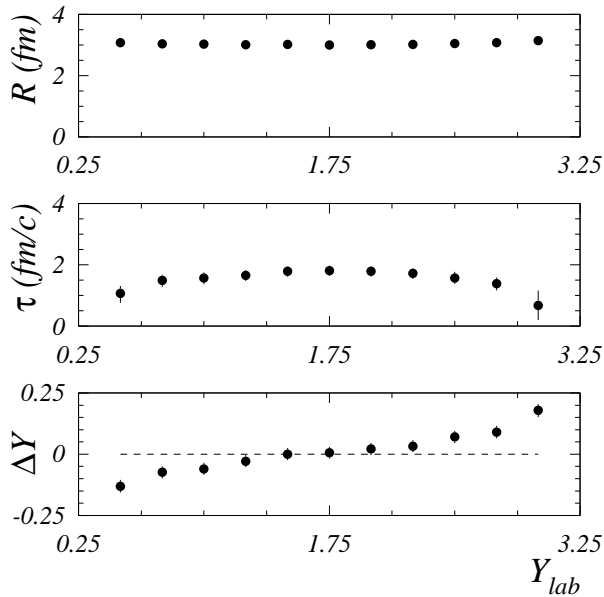}
 \caption{Systematic dependence of Yano-Koonin fit parameters on
 reference frame for Monte Carlo Gaussian sources, with $R=3$~fm,
 $\tau=2$~fm/c and a source velocity corresponding to $Y=1.75$.
 $\Delta Y$ refers to the difference between the fit rapidity and the
 rapidity of the Monte Carlo Gaussian source.}
\label{fig:yksrc}
\end{figure}

To study systematic effects of the Yano-Koonin parameterization a
Monte Carlo sample of pion events was generated for a 3~fm 
spherically symmetric Gaussian source with lifetime 2~fm/c at $Y=1.75$
in the lab.  These simulated pairs were then fit to Eq.~\ref{eq:yk}
while varying the fit frame (in which the q-variables were calculated)
from $Y=0.5$ to $Y=3.0$ in steps of 0.25 units of rapidity.
In all fits, the bin sizes were fixed at 20~MeV for \ql~and 15~MeV for
\Qinv~and \qzero.  Fig.~\ref{fig:yksrc} shows the results of this
study for the fitted radius, lifetime, and source frame.
There is a systematic shift in the source frame that is approximately
10\% of the difference between the fit frame and the actual source
frame.  In the simulation this difference is noticeable for
differences as small as 0.25 units in rapidity.  The change in source
frame is correlated with a reduction in the measured lifetime,
although $R$ is observed to be stable for the region tested.

\section{Coulomb Correction}
\label{sec:coulomb}

The Coulomb correction function used in this work is an analytic
approximation to the full Coulomb wave calculation.  The form of this
correction is given by, 
\begin{equation}
\label{eq:coulmdb}
G(\eta) \times \left[ 1 + \eta P5(s) \frac{(1 + F(s))}{1 + e^{-s^2}} \right]
\end{equation}
where $G$ is the standard Gamow correction for a point source,
and $P5$ is a fifth order polynomial in the dimensionless quantity
$s$,
\begin{eqnarray}
\label{eq:gamow}
G(\eta) & = & \frac{2 \pi \eta}{e^{2 \pi \eta} - 1 } \\
\eta    & = & 2 \mu \alpha / \Qinv \\
s       & = & \Rinv \Qinv / \hbar \\
P5(s)   & = & \frac{2}{\sqrt{\pi}} \min(\pi,8(s - 5s^3/9 + 52s^5/225))
\end{eqnarray}
where, $\mu$ is the reduced mass of the pion pair and $\alpha$ is the
fine structure constant.  The full form of the integral equation that
was solved is described elsewhere~\cite{e802qm96}, and the FORTRAN
code for the tabulated function $F(s)$ is available upon
request~\cite{coulnote}.

\begin{figure}[htb]
\includegraphics[width=3in]{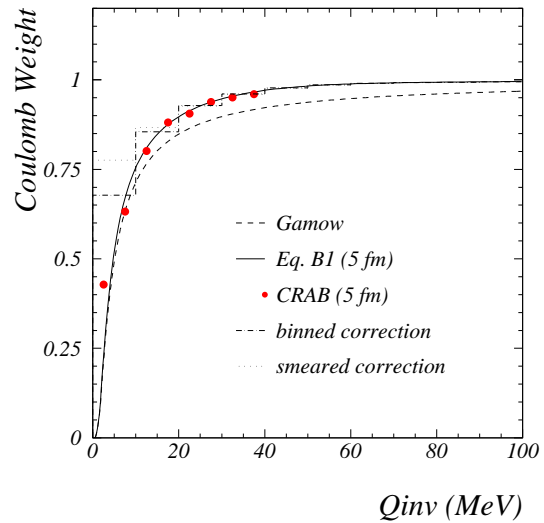}
\caption{Coulomb corrections for a point source (Gamow), a 5~fm
spherically symmetric Gaussian calculated with (CRAB) and
Eq.~\ref{eq:coulmdb}.  The dashed histogram shows the effect of
averaging Eq.~\ref{eq:coulmdb} over the Si+Au data (set 5), and the
dot-dashed histogram incorporates the effect of momentum resolution
smearing.}
\label{fig:coulomb}
\end{figure}

The accuracy of this analytic approximation (Eq.~\ref{eq:coulmdb}) is
demonstrated in Fig.~\ref{fig:coulomb} by comparing the correction
function for a 5~fm source to the Coulomb Wave integration calculated
by the Correlation After Burner code (CRAB)~\cite{crab}.
Fig.~\ref{fig:coulomb} shows a significant difference between the full
Coulomb correction and the and Gamow correction, but the fast
approximation shows good agreement with the calculation from CRAB.

The effect of resolution smearing was estimated by evaluating
Eq.~\ref{eq:coulmdb} for the event-mixed background, but applying the
Coulomb weight at the value of \Qinv~returned by a spectral response
function.  The form of the response function was parameterized in
momentum and polar angle spread according to Monte Carlo single
particle distributions~\cite{cia94}.  Fig.~\ref{fig:coulomb} shows the
Coulomb correction histogram --- weighted background divided by
background --- with and without the response function.  The effect of
the response function is evident only in the lowest bin (0-10~MeV/c),
and introduced a 1\% reduction in the 3D HBT radii and $\lambda$
values.  Given the small size of the effect, it was incorporated into
the overall systematic error estimate, but the correlation fits used
the unsmeared correction.

\begin{table}
\caption{Bertsch-Pratt fit parameters for full, half, and no Coulomb correction.}
\label{tab:coul}
\begin{ruledtabular}
\begin{tabular}{c||c|c|c}
Parameter  &  full  &  half  & none \\
\hline
$\lambda$  & 0.645$\pm$0.045 & 0.397$\pm$0.019 & 0.354$\pm$0.020 \\
\Rl        & 2.96$\pm$0.23   & 2.53$\pm$0.15   & 2.32$\pm$0.17   \\
\Rs        & 2.30$\pm$0.24   & 2.21$\pm$0.21   & 1.93$\pm$0.23   \\
\Ro        & 3.34$\pm$0.17   & 3.94$\pm$0.13   & 3.79$\pm$0.16   \\
\chisq/dof & 3266.3/3540     & 4270.6/4298     &  4284.8/4298    \\
\end{tabular}
\end{ruledtabular}
\end{table}
If the deviation of $\lambda$ from unity is caused by a substantial
fraction of pions from long lived resonances, it is
possible that the effect of the Coulomb interaction will also be reduced in
strength.  To understand the magnitude of this effect, the \siaupm\
data sets were fit with a half-strength Coulomb correction, and also
with no Coulomb correction.  The results are summarized in
Table~\ref{tab:coul}.  Variations in the radii are as large as 20\%
for \Rl~and \Ro, and $\lambda$ is greatly affected.  A thorough
investigation of effect of long lived resonances on the Coulomb
correction is left for further study.


\section{Tables of Correlation Radii}
\label{sec:tables}

Tables~\ref{tab:qinv} through~\ref{tab:bphi} show the complete set of
fitted radii used to study the system, centrality, and transverse mass
dependence for the \Qinv, Yano-Koonin, and Bertsch-Pratt fits of the
14.6~\AGeV~Si+Al, Si+Au, and 11.6~\AGeV~Au+Au data sets.  All $R$ and
$\tau$ parameters are given in units of fm and fm/c, respectively.

\begin{table*}[!]
\caption{\Qinv~fit parameters and \mmT~for data sets 1--9.}
\label{tab:qinv}
\begin{ruledtabular}
\begin{tabular}{c|c|c|c|c|c|c}
 System   &  Set  &  \mmT & \mmTs & $\lambda$ & \Rinv & \chisq/dof \\ \hline
\multirow{2}*{Si+Al}  & 1 & 0.348 & 0.327 & 0.491 $\pm$ 0.029 & 3.59 $\pm$ 0.16 &  32.1/27 \\
                      & 2 & 0.349 & 0.319 & 0.676 $\pm$ 0.040 & 4.42 $\pm$ 0.16 &  20.8/27 \\ \hline
\multirow{4}*{Si+Au}  & 3 & 0.345 & 0.323 & 0.487 $\pm$ 0.039 & 3.70 $\pm$ 0.23 &  46.6/27 \\
                      & 4 & 0.345 & 0.313 & 0.551 $\pm$ 0.035 & 4.47 $\pm$ 0.18 &  45.0/27 \\
                      & 5 & 0.342 & 0.306 & 0.511 $\pm$ 0.026 & 4.91 $\pm$ 0.15 &  39.2/27 \\
                      & 6 & 0.349 & 0.316 & 0.514 $\pm$ 0.041 & 4.54 $\pm$ 0.22 &  11.5/27 \\ \hline
\multirow{3}*{Au+Au}  & 7 & 0.411 & 0.361 & 0.441 $\pm$ 0.039 & 5.13 $\pm$ 0.28 &  24.1/27 \\
                      & 8 & 0.410 & 0.348 & 0.451 $\pm$ 0.051 & 6.55 $\pm$ 0.43 &  31.3/27 \\
                      & 9 & 0.420 & 0.358 & 0.428 $\pm$ 0.046 & 6.09 $\pm$ 0.38 &  24.8/26 \\
\end{tabular}		  
\end{ruledtabular}
\end{table*}
\begin{table*}[!]
\caption{Yano-Koonin fit parameters for data sets 1--9.}
\label{tab:fityk}
\begin{ruledtabular}
\begin{tabular}{c|c||c|c|c|c|c|c}
System & Set & \Yfit & $\lambda$ & R & $\tau$ & $\beta$ & $\chi^2/$dof \\ 
\hline
\multirow{2}*{Si+Al}  
 & 1   & 1.72 & 0.452 $\pm$ 0.024 & 1.74 $\pm$ 0.13 & 2.17 $\pm$ 0.19 & -0.06 $\pm$ 0.06 & 3770.7/4044      \\
 & 2   & 1.72 & 0.627 $\pm$ 0.035 & 2.75 $\pm$ 0.14 & 2.01 $\pm$ 0.38 & -0.08 $\pm$ 0.07 & 3901.9/3992      \\ 
\hline
\multirow{4}*{Si+Au}  
 & 3   & 1.72 & 0.467 $\pm$ 0.024 & 2.17 $\pm$ 0.11 & 2.66 $\pm$ 0.27 & -0.04 $\pm$ 0.05 & 3357.6/3605      \\
 & 4   & 1.72 & 0.503 $\pm$ 0.028 & 2.45 $\pm$ 0.16 & 2.70 $\pm$ 0.31 & -0.03 $\pm$ 0.06 & 4048.3/4156      \\
 & 5   & 1.50 & 0.499 $\pm$ 0.022 & 2.83 $\pm$ 0.09 & 3.01 $\pm$ 0.25 & -0.06 $\pm$ 0.05 & 4588.6/4844      \\
 & 6   & 1.50 & 0.515 $\pm$ 0.038 & 2.76 $\pm$ 0.20 & 3.10 $\pm$ 0.39 &  0.07 $\pm$ 0.07 & 3811.3/4055      \\  
\hline
\multirow{3}*{Au+Au}  
 & 7   & 1.45 & 0.496 $\pm$ 0.031 & 2.84 $\pm$ 0.14 & 3.86 $\pm$ 0.30 & 0.01 $\pm$ 0.07 & 4234.2/4407     \\
 & 8   & 1.45 & 0.454 $\pm$ 0.044 & 3.36 $\pm$ 0.27 & 4.48 $\pm$ 0.54 &-0.12 $\pm$ 0.08 & 4382.3/4503     \\
 & 9   & 1.45 & 0.530 $\pm$ 0.053 & 3.96 $\pm$ 0.25 & 4.23 $\pm$ 0.88 & 0.11 $\pm$ 0.10 & 4136.8/4486     \\
\end{tabular}
\end{ruledtabular}
\end{table*}
\begin{table*}[!]
\caption{Bertsch-Pratt fit parameters for data sets 1--9.  Values for
$\tau$ are calculated from Eq.~\ref{eq:bptau} with \Rs--\Ro~covariances used in
error propagation.}
\label{tab:bp}
\begin{ruledtabular}
\begin{tabular}{c|c||c|c|c|c|c|c}
System & Set & $\lambda$ & \Rl & \Rs & \Ro & $\tau$          & $\chi^2$/dof \\
\hline
\multirow{2}*{Si+Al} 
 & 1 & 0.437 $\pm$ 0.024 & 1.76 $\pm$ 0.17 & 1.05 $\pm$ 0.31 & 2.61 $\pm$ 0.13 & 2.71$\pm$0.14 & 3507.3/3594  \\
 & 2 & 0.645 $\pm$ 0.045 & 2.96 $\pm$ 0.23 & 2.30 $\pm$ 0.24 & 3.34 $\pm$ 0.17 & 3.13$\pm$0.26 & 3266.3/3540  \\ 
\hline
\multirow{4}*{Si+Au} 
 & 3 & 0.452 $\pm$ 0.034 & 2.25 $\pm$ 0.22 & 1.63 $\pm$ 0.43 & 3.21 $\pm$ 0.18 & 3.06$\pm$0.10 & 3053.2/3184  \\
 & 4 & 0.488 $\pm$ 0.036 & 2.29 $\pm$ 0.21 & 2.12 $\pm$ 0.30 & 3.43 $\pm$ 0.18 & 2.75$\pm$0.20 & 3382.1/3636  \\
 & 5 & 0.503 $\pm$ 0.021 & 2.82 $\pm$ 0.14 & 2.65 $\pm$ 0.18 & 3.89 $\pm$ 0.13 & 3.32$\pm$0.20 & 4274.9/4298  \\
 & 6 & 0.517 $\pm$ 0.054 & 2.57 $\pm$ 0.27 & 2.56 $\pm$ 0.39 & 3.97 $\pm$ 0.28 & 3.52$\pm$0.31 & 3544.5/3672  \\ 
\hline
\multirow{3}*{Au+Au} 
 & 7 & 0.506 $\pm$ 0.037 & 2.62 $\pm$ 0.23 & 3.00 $\pm$ 0.33 & 4.66 $\pm$ 0.26 & 3.95$\pm$0.37 & 2942.1/3114  \\
 & 8 & 0.480 $\pm$ 0.043 & 3.25 $\pm$ 0.32 & 3.84 $\pm$ 0.46 & 5.35 $\pm$ 0.33 & 4.13$\pm$0.56 & 3070.2/3236  \\
 & 9 & 0.536 $\pm$ 0.047 & 3.92 $\pm$ 0.37 & 3.79 $\pm$ 0.45 & 5.60 $\pm$ 0.43 & 4.55$\pm$0.61 & 2694.4/3095  \\
\end{tabular}
\end{ruledtabular}
\end{table*}
\begin{table*}[!]
\caption{\Qinv~and Bertsch-Pratt fit parameters with \Rlo\ cross-term
(see Eq.~\ref{eq:xtrm}) for data sets 1--9.}
\label{tab:rlo}
\begin{ruledtabular}
\begin{tabular}{c|c||c|c|c|c|c|c}
System & Set & $\lambda$ & \Rl & \Rs & \Ro & \Rlo & $\chi^2/$dof \\
\hline
\multirow{2}*{Si+Al} 
 & 1 & 0.430 $\pm$ 0.023 & 1.65 $\pm$ 0.17 & 0.97 $\pm$ 0.31 & 2.59 $\pm$ 0.13 & 0.58 $\pm$ 0.36 & 3441.8/3604 \\
 & 2 & 0.618 $\pm$ 0.034 & 2.71 $\pm$ 0.19 & 2.20 $\pm$ 0.23 & 3.29 $\pm$ 0.15 & 0.51 $\pm$ 0.69 & 3363.7/3533 \\ 
\hline
\multirow{4}*{Si+Au} 
 & 3 & 0.426 $\pm$ 0.034 & 2.03 $\pm$ 0.23 & 1.26 $\pm$ 0.49 & 3.15 $\pm$ 0.24 & 0.74 $\pm$ 0.82 & 3260.6/3200 \\
 & 4 & 0.479 $\pm$ 0.028 & 2.16 $\pm$ 0.18 & 2.01 $\pm$ 0.26 & 3.41 $\pm$ 0.16 & 0.68 $\pm$ 0.68 & 3303.2/3643 \\
 & 5 & 0.496 $\pm$ 0.021 & 2.75 $\pm$ 0.14 & 2.59 $\pm$ 0.18 & 3.86 $\pm$ 0.13 & 2.59 $\pm$ 0.75 & 4126.7/4293 \\
 & 6 & 0.496 $\pm$ 0.037 & 2.47 $\pm$ 0.24 & 2.47 $\pm$ 0.31 & 3.93 $\pm$ 0.24 & 0.50 $\pm$ 0.94 & 3498.9/3621 \\ 
\hline
\multirow{3}*{Au+Au} 
 & 7 & 0.499 $\pm$ 0.036 & 2.55 $\pm$ 0.23 & 2.90 $\pm$ 0.32 & 4.66 $\pm$ 0.26 & -1.34 $\pm$ 1.29 & 2919.7/3170      \\
 & 8 & 0.464 $\pm$ 0.040 & 3.18 $\pm$ 0.37 & 3.50 $\pm$ 0.45 & 5.37 $\pm$ 0.22 &  4.58 $\pm$ 2.37 & 3018.2/3264      \\ 
 & 9 & 0.582 $\pm$ 5.033 & 4.49 $\pm$ 1.15 & 3.71 $\pm$ 1.00 & 5.74 $\pm$ 1.09 & -3.10 $\pm$ 3.08 & 3018.2/3264      \\
\end{tabular}
\end{ruledtabular}
\end{table*}
\begin{table*}[!]
\caption{LCMS frame Bertsch-Pratt fit parameters for the for data sets 1--9.}
\label{tab:bplcms}
\begin{ruledtabular}
\begin{tabular}{c|c||c|c|c|c|c}
System & Set & $\lambda$ & \Rl & \Rs & \Ro & $\chi^2$/dof \\
\hline
\multirow{2}*{Si+Al} 
 & 1 & 0.359 $\pm$ 0.022 & 1.42 $\pm$ 0.22 & 0.00 $\pm$ 0.48 & 2.53 $\pm$ 0.14 & 3365.7/3503      \\
 & 2 & 0.531 $\pm$ 0.041 & 2.68 $\pm$ 0.25 & 2.00 $\pm$ 0.27 & 3.35 $\pm$ 0.19 & 3163.2/3478      \\
\hline
\multirow{4}*{Si+Au} 
 & 3 & 0.359 $\pm$ 0.038 & 1.96 $\pm$ 0.30 & 0.90 $\pm$ 0.82 & 3.12 $\pm$ 0.26 & 3113.5/3107      \\
 & 4 & 0.392 $\pm$ 0.028 & 1.97 $\pm$ 0.22 & 1.57 $\pm$ 0.37 & 3.41 $\pm$ 0.17 & 3206.7/3551      \\
 & 5 & 0.398 $\pm$ 0.023 & 2.66 $\pm$ 0.19 & 2.23 $\pm$ 0.23 & 4.05 $\pm$ 0.17 & 3835.7/4199      \\
 & 6 & 0.413 $\pm$ 0.042 & 2.43 $\pm$ 0.31 & 2.11 $\pm$ 0.46 & 4.17 $\pm$ 0.27 & 3261.9/3447      \\
\hline
\multirow{3}*{Au+Au} 
 & 7 & 0.399 $\pm$ 0.043 & 2.28 $\pm$ 0.30 & 2.54 $\pm$ 0.46 & 4.77 $\pm$ 0.38 & 3002.2/3084      \\
 & 8 & 0.347 $\pm$ 0.037 & 2.75 $\pm$ 0.36 & 3.11 $\pm$ 0.57 & 5.49 $\pm$ 0.39 & 3034.3/3188      \\
 & 9 & 0.392 $\pm$ 0.042 & 3.49 $\pm$ 0.44 & 3.05 $\pm$ 0.56 & 6.05 $\pm$ 0.58 & 2714.5/3033      \\
\end{tabular}
\end{ruledtabular}
\end{table*}

\begin{table*}
\caption{\Qinv~and Bertsch-Pratt fit parameters for \mT-dependent data sets a--h.}
\label{tab:bpkt}
\begin{ruledtabular}
\begin{tabular}{c|c||c|c||c|c|c|c|c|c}
System & Set & \Rinv & \mmTs & $\lambda$ & \Rl & \Rs & \Ro & $\tau$ & \chisq/dof \\ 
\hline
\multirow{5}*{Si+Au}
 & a & 5.03 $\pm$ 0.21 & 0.248 & 0.560 $\pm$ 0.035 & 3.52 $\pm$ 0.25 & 3.25 $\pm$ 0.28 & 4.33 $\pm$ 0.22 & 3.32$\pm$0.35 & 1834.6/1844 \\
 & b & 4.69 $\pm$ 0.29 & 0.266 & 0.577 $\pm$ 0.059 & 3.10 $\pm$ 0.31 & 3.15 $\pm$ 0.45 & 4.01 $\pm$ 0.30 & 2.75$\pm$0.54 & 2036.9/2083 \\
 & c & 4.84 $\pm$ 0.29 & 0.333 & 0.530 $\pm$ 0.064 & 2.81 $\pm$ 0.29 & 2.69 $\pm$ 0.34 & 3.53 $\pm$ 0.25 & 2.49$\pm$0.40 & 2796.6/3007 \\
 & d & 3.94 $\pm$ 0.34 & 0.408 & 0.512 $\pm$ 0.075 & 2.11 $\pm$ 0.28 & 2.18 $\pm$ 0.44 & 3.84 $\pm$ 0.46 & 3.32$\pm$0.53 & 3705.3/3657 \\
 & e & 3.93 $\pm$ 0.25 & 0.441 & 0.628 $\pm$ 0.058 & 2.30 $\pm$ 0.17 & 2.54 $\pm$ 0.25 & 3.05 $\pm$ 0.20 & 1.78$\pm$0.49 & 4296.6/4230 \\
\hline
\multirow{3}*{Au+Au} 
 & f & 6.05 $\pm$ 0.38 & 0.294 & 0.544 $\pm$ 0.055 & 3.62 $\pm$ 0.38 & 4.90 $\pm$ 0.51 & 5.57 $\pm$ 0.41 & 2.92$\pm$0.93 & 1261.7/1215 \\
 & g & 6.68 $\pm$ 0.85 & 0.400 & 0.479 $\pm$ 0.073 & 3.05 $\pm$ 0.42 & 3.11 $\pm$ 0.54 & 4.39 $\pm$ 0.44 & 3.32$\pm$0.64 & 1960.6/2010 \\
 & h & 5.76 $\pm$ 1.09 & 0.539 & 0.508 $\pm$ 0.101 & 2.36 $\pm$ 0.34 & 3.73 $\pm$ 0.57 & 4.21 $\pm$ 0.54 & 2.10$\pm$1.44 & 3541.3/3693 \\
\end{tabular}
\end{ruledtabular}
\end{table*}
\begin{table*}[!]
\caption{\Qinv~and Bertsch-Pratt fit parameters for low-\mT~data sets 1--9.}
\label{tab:bplo}
\begin{ruledtabular}
\begin{tabular}{c|c|c||c|c||c|c|c|c|c}
System & Set & \mmTs  & \Rinv & $\lambda$ & \Rl & \Rs & \Ro & $\chi^2/$dof \\
\hline
\multirow{2}*{Si+Al} 
& 1 & 0.277 & 3.38 $\pm$ 0.19 & 0.502 $\pm$ 0.038 & 2.34 $\pm$ 0.25 & 2.15 $\pm$ 0.32 & 2.75 $\pm$ 0.18 & 1867.9/1849 \\
& 2 & 0.271 & 4.37 $\pm$ 0.20 & 0.685 $\pm$ 0.044 & 3.36 $\pm$ 0.25 & 2.58 $\pm$ 0.28 & 3.46 $\pm$ 0.20 & 1789.6/1854 \\
 \hline
 \multirow{4}*{Si+Au} 
& 3 & 0.270 & 4.23 $\pm$ 0.31 & 0.500 $\pm$ 0.070 & 2.59 $\pm$ 0.45 & 2.30 $\pm$ 0.63 & 3.39 $\pm$ 0.36 & 1589.0/1699 \\
& 4 & 0.269 & 4.45 $\pm$ 0.21 & 0.569 $\pm$ 0.042 & 2.90 $\pm$ 0.27 & 2.86 $\pm$ 0.33 & 3.49 $\pm$ 0.22 & 1796.6/1859 \\
& 5 & 0.264 & 5.01 $\pm$ 0.17 & 0.539 $\pm$ 0.027 & 3.02 $\pm$ 0.18 & 3.18 $\pm$ 0.22 & 4.05 $\pm$ 0.17 & 2939.3/2981 \\
& 6 & 0.268 & 4.72 $\pm$ 0.29 & 0.576 $\pm$ 0.046 & 3.07 $\pm$ 0.28 & 3.08 $\pm$ 0.38 & 4.03 $\pm$ 0.26 & 2105.1/2151 \\
 \hline
 \multirow{2}*{Au+Au} 
& 7 & 0.309 & 5.20 $\pm$ 0.33 & 0.522 $\pm$ 0.051 & 2.87 $\pm$ 0.32 & 3.36 $\pm$ 0.50 & 4.52 $\pm$ 0.35 & 1305.1/1427 \\
&8+9& 0.308 & 6.08 $\pm$ 0.37 & 0.542 $\pm$ 0.041 & 3.61 $\pm$ 0.30 & 4.55 $\pm$ 0.42 & 5.41 $\pm$ 0.33 & 1640.1/1532 \\
\end{tabular}
\end{ruledtabular}
\end{table*}
\begin{table*}[!]
\caption{\Qinv~and Bertsch-Pratt fit parameters for high-\mT~data sets 1--9.}
\label{tab:bphi}
\begin{ruledtabular}
\begin{tabular}{c|c|c||c|c||c|c|c|c|c}
System & Set & \mmTs  & \Rinv & $\lambda$ & \Rl & \Rs & \Ro & $\chi^2/$dof \\
\hline
\multirow{2}*{Si+Al} 
& 1 & 0.410 & 3.58 $\pm$ 0.26 & 0.429 $\pm$ 0.039 & 1.37 $\pm$ 0.41 & 0.13 $\pm$ 0.78 & 2.31 $\pm$ 0.35 & 3491.7/3590 \\
& 2 & 0.411 & 4.28 $\pm$ 0.24 & 0.760 $\pm$ 0.069 & 2.65 $\pm$ 0.25 & 2.41 $\pm$ 0.26 & 3.08 $\pm$ 0.21 & 3428.1/3534 \\
 \hline
 \multirow{4}*{Si+Au} 
& 3 & 0.411 & 2.98 $\pm$ 0.23 & 0.472 $\pm$ 0.072 & 2.26 $\pm$ 0.27 & 1.38 $\pm$ 0.57 & 2.97 $\pm$ 0.27 & 3187.4/3165 \\
& 4 & 0.411 & 3.93 $\pm$ 0.31 & 0.519 $\pm$ 0.052 & 1.97 $\pm$ 0.19 & 1.82 $\pm$ 0.31 & 3.17 $\pm$ 0.19 & 3523.8/3633 \\
& 5 & 0.411 & 4.31 $\pm$ 0.24 & 0.573 $\pm$ 0.064 & 2.15 $\pm$ 0.20 & 2.36 $\pm$ 0.27 & 3.15 $\pm$ 0.21 & 4840.6/4986 \\
& 6 & 0.412 & 3.90 $\pm$ 0.34 & 0.495 $\pm$ 0.079 & 2.14 $\pm$ 0.26 & 2.14 $\pm$ 0.49 & 3.72 $\pm$ 0.42 & 3669.6/3654 \\
 \hline
 \multirow{2}*{Au+Au} 
& 7 & 0.485 & 4.24 $\pm$ 0.47 & 0.644 $\pm$ 0.089 & 2.68 $\pm$ 0.34 & 3.02 $\pm$ 0.40 & 4.72 $\pm$ 0.52 & 3174.5/3113 \\
&8+9& 0.485 & 6.44 $\pm$ 0.85 & 0.465 $\pm$ 0.056 & 2.34 $\pm$ 0.28 & 3.41 $\pm$ 0.39 & 4.01 $\pm$ 0.35 & 3581.9/3701 \\
\end{tabular}
\end{ruledtabular}
\end{table*}

\printtables


\end{document}